\documentstyle[preprint,aps,prd,psfig]{revtex}

\newcommand{\la}{\langle}
\newcommand{\ra}{\rangle}

\newcommand{\AmS}{{\protect\the\textfont2
  A\kern-.1667em\lower.5ex\hbox{M}\kern-.125emS}}
\newcommand{\be}{\begin{equation}}
\newcommand{\ee}{\end{equation}}
\newcommand{\ben}{\begin{eqnarray}}
\newcommand{\een}{\end{eqnarray}}
\newcommand{\nn}{\nonumber}
\newcommand{\tr}{{\rm Tr}}
\newcommand{\slas}[2]{{{#1}\hspace{-5pt}{/}}_{#2}}
\newcommand{\slal}[2]{{{#1}\hspace{-5pt}{/}}_{#2}}
\def\simgt{\rlap{\lower 3.5 pt\hbox{$\mathchar \sim$}}\raise 1pt \hbox {$>$}}
\def\simlt{\rlap{\lower 3.5 pt\hbox{$\mathchar \sim$}}\raise 1pt \hbox {$<$}}

\begin{document}
\draft
\title{Nucleon Decay Matrix Elements from Lattice QCD}

\author{JLQCD Collaboration \\[1mm]
        S.~Aoki$^{\rm a}$,
        M.~Fukugita$^{\rm b}$,
        S.~Hashimoto$^{\rm c}$,
        K-I.~Ishikawa$^{\rm d}$,
        N.~Ishizuka$^{\rm a,e}$
        Y.~Iwasaki$^{\rm a,e}$,
        K.~Kanaya$^{\rm a,e}$,
        T.~Kaneda$^{\rm a}$,
        S.~Kaya$^{\rm d}$,
        Y.~Kuramashi$^{\rm f}$
\thanks{On leave from Institute of Particle and Nuclear Studies,
        High Energy Accelerator Research Organization(KEK),
        Tsukuba, Ibaraki 305-0801, Japan},
        M.~Okawa$^{\rm d}$,
        T.~Onogi$^{\rm g}$,
        S.~Tominaga$^{\rm d}$,
        N.~Tsutsui$^{\rm g}$,
        A.~Ukawa$^{\rm a,e}$,
        N.~Yamada$^{\rm g}$,
        T.~Yoshi\'e$^{\rm a,e}$
       }
\address{$^a$Institute of Physics,
             University of Tsukuba,\\
             Tsukuba, Ibaraki 305-8571, Japan\\
         $^b$Institute for Cosmic Ray Research,
             University of Tokyo,\\
             Tanashi, Tokyo 188-8502, Japan\\
         $^c$Computing Research Center, 
             High Energy Accelerator Research Organization(KEK),\\
             Tsukuba, Ibaraki 305-0801, Japan\\
         $^d$Institute of Particle and Nuclear Studies, 
             High Energy Accelerator Research Organization(KEK),\\ 
             Tsukuba, Ibaraki 305-0801, Japan\\
         $^e$Center for Computational Physics,
             University of Tsukuba,\\
             Tsukuba, Ibaraki 305-8577, Japan\\
         $^f$Department of Physics,
             Washington University,\\
             St. Louis, Missouri 63130, USA\\
         $^g$Department of Physics,
             Hiroshima University,\\
             Higashi-Hiroshima, Hiroshima 739-8526, Japan
        }

\date{\today}

\maketitle

\newpage
\begin{abstract}

We present a model-independent calculation of hadron matrix elements
for all dimension-six operators associated with baryon number 
violating processes using lattice QCD. 
The calculation is performed with the
Wilson quark action in the quenched approximation at 
$\beta=6/g^2=6.0$ on a $28^2\times 48\times 80$ lattice.
Our results cover all the matrix elements required to estimate
the partial lifetimes of (proton,neutron)$\rightarrow$($\pi,K,\eta$)
+(${\bar \nu},e^+,\mu^+$) decay modes.
We point out the necessity of disentangling two form factors that contribute
to the matrix element; previous calculations did not make the separation, 
which led to an underestimate of the physical matrix elements.
With a correct separation, we find that the matrix elements
have values $3-5$ times larger than the smallest
estimates employed in phenomenological analyses of the nucleon
decays, which could give strong constraints on
several GUT models.
We also find that the values of the matrix elements are comparable
with the tree-level predictions of chiral lagrangian. 

\end{abstract}

\section{Introduction}

Nucleon decay is one of the most exciting predictions of
grand unified theories (GUTs) regardless of the existence
of supersymmetry (SUSY). 
Although none of the decay modes have been detected up to now,
experimental efforts over the years have pushed 
the lower limit on the partial lifetimes of the nucleon.
Moreover, an improvement by an order of magnitude is expected
from the Super-Kamiokande experiment,
which can give a strong constraint on (SUSY-)GUTs.
On the other hand, theoretical predictions of the nucleon 
partial lifetimes suffer from various uncertainties.
One of the main sources of uncertainties is found in the evaluation
of the hadron matrix elements for the nucleon decays  
$\langle PS | {\cal O}^{\slal{B}{}} |N\rangle$, where
$PS$ and  $N$ denote the pseudoscalar
meson and the nucleon, and ${\cal O}^{\slal{B}{}}$
is the baryon number violating operator that appears in the 
low-energy effective lagrangian of (SUSY-)GUTs.
The matrix elements have been estimated by
employing various QCD models. Their results, however,
scatter over the range whose minimum and maximum values differ by 
a factor of ten\cite{model_wf}.
Therefore, a precise determination of the nucleon decay matrix elements
from the first principles using lattice QCD is of extreme importance.

In lattice QCD the pioneering studies for the nucleon decay 
matrix elements\cite{hara,bowler} attempted to estimate the matrix element
$\langle \pi^0 | {\cal O}^{\slal{B}{}} |p\rangle$,
which is relevant to the dominant decay mode $p\rightarrow\pi^0+e^+$
in the minimal SU(5) GUT,
from the matrix element $\langle 0 | {\cal O}^{\slal{B}{}} |p\rangle$
with the aid of chiral perturbation theory.
This was followed by a direct measurement of 
$\langle \pi^0 | {\cal O}^{\slal{B}{}} |p\rangle$
with the use of the three-point functions\cite{gavela}, which 
showed an unexpectedly large discrepancy
between these two methods:
the direct method yielded a value of the matrix element 
two or three times smaller than the value obtained by the indirect method.
Recently we have revisited this old problem and confirmed\cite{jlqcd_98}
the peculiar feature when one follows the methods employed in the earlier
work\cite{gavela}.
 
In this paper we report results of our effort to advance the lattice QCD 
calculation of the nucleon decay matrix elements in several directions. 
We point out that there are two form factors that contribute 
to the matrix element $\langle \pi^0 | {\cal O}^{\slal{B}{}} |p\rangle$
for general lepton momentum.  
While only one of the form factors is relevant 
for the physical amplitudes 
as the other form factor contribution is annulled by the negligibly small 
lepton mass, the two contributions have to be disentangled in the lattice
QCD calculation. 
This explains the discrepancy between the direct and indirect 
estimations of the proton decay matrix element found in the 
previous studies\cite{gavela,jlqcd_98} 
where the separation was not made. 

Another important feature of our calculation is 
model independence.
All dimension-six operators associated with baryon number 
violating processes 
are classified into four types under the requirement of
SU(3)$\times$SU(2)$\times$U(1) invariance 
at low-energy scales\cite{op_6,op_4}. 
If one specifies the decay processes of interest,
namely the processes among (proton,neutron)$\rightarrow$($\pi,K,\eta$)
+(${\bar \nu},e^+,\mu^+$), we can list a complete set of
independent matrix elements in QCD, and we calculate 
all the matrix elements.  

Other advances, which are more technical but essential for precise 
calculation, are the following two points: 
(i) flavor SU(3) breaking effect in the process with the $K$ meson 
in the final state is correctly taken into account by setting
the strange quark mass non-degenerate with the up and down 
quark mass, (ii) two spatial momenta are injected
to investigate the $q^2$ dependence of the matrix elements, where
$q$ is the four-momentum transfer between the nucleon and the 
pseudoscalar meson.

This paper is organized as follows.
In Sec.~\ref{sec:method} we formulate our calculational
method of the nucleon decay matrix elements.
The complete set of the independent matrix elements
is also presented. In Sec.~\ref{sec:chpt} we briefly review
the chiral lagrangian for the baryon number violating
interactions and enumerate its tree-level predictions. 
Section~\ref{sec:parameter} 
contains the simulation parameters and technical details.
Results for the matrix element $\langle 0 | {\cal O}^{\slal{B}{}} |p\rangle$
are given in Sec.~\ref{sec:ab}.
In Sec.~\ref{sec:ndecay} we present the results
for the nucleon decay matrix elements obtained by the direct method 
and compare them with the tree-level predictions of the chiral lagrangian.
We also discuss the soft pion limit of the matrix elements.
Our conclusions are summarized in Sec.~\ref{sec:conclusion}.

\section{Formulation of the method}
\label{sec:method}
\subsection{Independent matrix elements for nucleon decays}

One of the most important features in the study of 
the baryon number violating processes is that 
the low energy effective theory is described
in terms of  SU(3)$\times$SU(2)$\times$U(1) gauge symmetry
based on the strong and the electroweak interactions,
which enables us to make a model independent analysis.
Our interest is focused on the dimension-six operators
which are the lowest dimensional operators 
in the low energy effective Hamiltonian:
operators associated with the baryon number violating processes
must contain at least three quark fields to form SU(3) color
singlet, and then an additional lepton field is
required to construct a Lorentz scalar.
Higher-dimensional operators 
are suppressed by inverse powers of heavy particle mass
that is characterized by the theory beyond the standard model.

All dimension-six operators are classified
into the four types under the
requirement of SU(3)$\times$SU(2)$\times$U(1) invariance 
\cite{op_6,op_4}:
\ben
{\cal O}^{(1)}_{abcd}&=&
({\bar D}^c_{i aR}U_{j bR})
({\bar Q}^c_{\alpha k cL}L_{\beta dL})
\epsilon_{ijk}\epsilon_{\alpha\beta},  
\label{eq:op_1}\\
{\cal O}^{(2)}_{abcd}&=&
({\bar Q}^c_{\alpha i aL}Q_{\beta j bL})
({\bar U}^c_{k cR}L_{dR})
\epsilon_{ijk}\epsilon_{\alpha\beta},  
\label{eq:op_2}\\
{\cal O}^{(3)}_{abcd}&=&
({\bar Q}^c_{\alpha i aL}Q_{\beta j bL})
({\bar Q}^c_{\gamma k cL}L_{\delta dL})
\epsilon_{ijk}\epsilon_{\alpha\delta}\epsilon_{\beta\gamma}, 
\label{eq:op_3}\\ 
{\cal O}^{(4)}_{abcd}&=&
({\bar D}^c_{i aR}U_{j bR})
({\bar U}^c_{k cR}L_{dR})
\epsilon_{ijk},
\label{eq:op_4}
\een 
where ${\bar \psi}^c=\psi^T C$ 
with $C=\gamma_4\gamma_2$ the charge conjugation matrix;
$i$, $j$ and $k$ are SU(3) color indices; 
$\alpha$, $\beta$, $\gamma$ and $\delta$ are SU(2) indices; 
$a$, $b$, $c$ and 
$d$ are generation indices; $L_L$ and $Q_L$ are generic lepton and quark
SU(2) doublets with the left-handed projection $P_L=(1-\gamma_5)/2$;
$L_R$, $U_R$, and $D_R$ are generic charged lepton and quark SU(2)
singlets with the right-handed projection $P_R=(1+\gamma_5)/2$.
Fierz transformations are used to eliminate all vector
and tensor Dirac structures in eqs.~(\ref{eq:op_1})$-$(\ref{eq:op_4}).

The operators relevant to non-strange final states are\cite{chpt_gut}
\ben
{\cal O}^{(1)}_{d}&=&
({\bar d}^c_{i R}u_{j R})
({\bar u}^c_{k L}e_{dL}-{\bar d}^c_{k L}\nu_{dL})
\epsilon_{ijk}, 
\label{eq:op_ns_1}\\
{\cal O}^{(2)}_{d}&=&
({\bar d}^c_{i L}u_{j L})
({\bar u}^c_{k R}e_{dR})
\epsilon_{ijk}, 
\label{eq:op_ns_2}\\
{\cal O}^{(3)}_{d}&=&
({\bar d}^c_{i L}u_{j L})
({\bar u}^c_{k L}e_{dL}-{\bar d}^c_{k L}\nu_{dL})
\epsilon_{ijk}, 
\label{eq:op_ns_3}\\
{\cal O}^{(4)}_{d}&=&
({\bar d}^c_{i R}u_{j R})
({\bar u}^c_{k R}e_{dR})
\epsilon_{ijk}.
\label{eq:op_ns_4}
\een
We can also list the operators relevant 
to strange final states\cite{chpt_gut}:
\ben
{\tilde {\cal O}}^{(1)}_{d}&=&
({\bar s}^c_{i R}u_{j R})
({\bar u}^c_{k L}e_{dL}-{\bar d}^c_{k L}\nu_{dL})
\epsilon_{ijk}, 
\label{eq:op_s_1}\\
{\tilde {\cal O}}^{(2)}_{d}&=&
({\bar s}^c_{i L}u_{j L})
({\bar u}^c_{k R}e_{dR})
\epsilon_{ijk}, 
\label{eq:op_s_2}\\
{\tilde {\cal O}}^{(3)}_{d}&=&
({\bar s}^c_{i L}u_{j L})
({\bar u}^c_{k L}e_{dL}-{\bar d}^c_{k L}\nu_{dL})
\epsilon_{ijk}, 
\label{eq:op_s_3}\\
{\tilde {\cal O}}^{(4)}_{d}&=&
({\bar s}^c_{i R}u_{j R})
({\bar u}^c_{k R}e_{dR})
\epsilon_{ijk},
\label{eq:op_s_4}\\
{\tilde {\cal O}}^{(5)}_{d}&=&
({\bar d}^c_{i R}u_{j R})
({\bar s}^c_{k L}\nu_{dL})
\epsilon_{ijk},
\label{eq:op_s_5}\\
{\tilde {\cal O}}^{(6)}_{d}&=&
({\bar d}^c_{i L}u_{j L})
({\bar s}^c_{k L}\nu_{dL})
\epsilon_{ijk},
\label{eq:op_s_6}
\een  
where $d$ denote the generation;
$e_1=e$, $e_2=\mu$, $\nu_1=\nu_e$ and $\nu_2=\nu_\mu$.

We are interested in the decay processes from the nucleon to one
pseudoscalar meson: (proton,neutron)$\rightarrow$($\pi,K,\eta$)
+(${\bar \nu},e^+,\mu^+$). For these decay modes
we can list the complete set of independent matrix elements
in QCD employing the operators of eqs.~(\ref{eq:op_ns_1})$-$(\ref{eq:op_s_6}):
\ben
&&\langle \pi^0|\epsilon_{ijk}
({u^i}^T CP_{R,L}d^j) P_L u^k|p\rangle, 
\label{eq:indme_1}\\ 
&&\langle \pi^+|\epsilon_{ijk}
({u^i}^T CP_{R,L}d^j) P_L d^k|p\rangle, 
\label{eq:indme_2}\\
&&\langle K^0|\epsilon_{ijk}
({u^i}^T CP_{R,L}s^j) P_L u^k|p\rangle, 
\label{eq:indme_3}\\
&&\langle K^+|\epsilon_{ijk}
({u^i}^T CP_{R,L}s^j) P_L d^k|p\rangle, 
\label{eq:indme_4}\\
&&\langle K^+|\epsilon_{ijk}
({u^i}^T CP_{R,L}d^j) P_L s^k|p\rangle, 
\label{eq:indme_5}\\
&&\langle K^0|\epsilon_{ijk}
({u^i}^T CP_{R,L}s^j) P_L d^k|n\rangle, 
\label{eq:indme_6}\\
&&\langle \eta |\epsilon_{ijk}
({u^i}^T CP_{R,L}d^j) P_L u^k|p\rangle,
\label{eq:indme_7} 
\een
where we assume SU(2) isospin symmetry $m_u=m_d$ and
use the relations
\ben
\la PS | {\cal O}_{LR} |N \ra&=&\la PS | {\cal O}_{RL} |N \ra,\\
\la PS | {\cal O}_{RR} |N \ra&=&\la PS | {\cal O}_{LL} |N \ra,
\een
due to the parity invariance.
All we have to calculate in lattice QCD are these 14 matrix elements.
Other matrix elements are obtained through
the exchange of the up and down quarks, under which
the nucleon and PS meson states transform as
\ben
&&|p\ra \rightarrow -|n\ra,\;
|n\ra \rightarrow -|p\ra, \\
&&\la \pi^+| \rightarrow \la \pi^{-}|,\;
\la \pi^0| \rightarrow -\la \pi^0|,\;
\la \pi^{-}| \rightarrow \la \pi^+|, \\
&&\la K^+| \rightarrow \la K^0|,\; 
\la K^0| \rightarrow \la K^+|, \\
&&\la \eta | \rightarrow \la \eta |,
\een
where there is no decay mode with the ${\bar K}^0$ or $K^{-}$
final state.

\subsection{Form factors in nucleon decay matrix elements}
\label{subsec:ff}

Under the requirement of Lorentz and parity invariance,
the matrix elements 
between the nucleon($N$) and
the pseudoscalar(PS) meson in eqs.~(\ref{eq:indme_1})$-$(\ref{eq:indme_7}) 
can have two form factors:
\be
\langle PS(\vec{p})|{\cal O}_L^{\slal{B}{}}|N^{(s)}(\vec{k})\rangle
=P_L\left(W_0(q^2)-W_q(q^2) i{\slas{q}{}}\right) u^{(s)},
\label{eq:ff}
\ee
where ${\cal O}^{\slal{B}{}}_L$ represents the three-quark operator
projected to the left-handed chiral state, 
$u^{(s)}$ denotes the Dirac spinor for nucleon  
with either the up ($s=1$) or down ($s=2$) spin state ,
and $q^2$ is the momentum squared of the out-going antilepton.
The contribution of the $W_q$ term in  eq.(\ref{eq:ff}) 
is negligible in the physical decay amplitude, because its contribution 
is of the order of the lepton mass $m_l$ 
after the multiplication with antilepton spinor. 
However, since the relative magnitude of the two form factors $W_0$ and 
$W_q$ is {\it a priori} not known, we have to disentangle these 
two form factors in the lattice QCD calculation. 
Hereafter we refer to $W_0$ and $W_q$ as relevant 
and irrelevant form factor respectively.

In the lattice calculation, $\vec{k}=\vec{0}$ is chosen for 
the nucleon spatial momentum and $\vec{p}=\vec{k}-\vec{q}\neq \vec{0}$ for
the PS meson.
In this case the Dirac structure of the right hand side 
in eq.~(\ref{eq:ff}) is given by
\ben
\left( W_0 - W_q i{\slas{q}{}}\right) u^{(s)}&=&
\left(
\begin{array}{cc}
     W_0 - iq_4 W_q       
&   -W_q {\vec q}\cdot\vec{\sigma}    \\
     W_q {\vec q}\cdot\vec{\sigma}  
&    W_0 + i q_4 W_q
\end{array}
\right) u^{(s)}
\nn \\
&=&
\left(
\begin{array}{cc}
    W_0 + (m_N-\sqrt{m_{PS}^2+{\vec p}^2}) W_q  
&   W_q {\vec p}\cdot\vec{\sigma}   \\
   -W_q {\vec p}\cdot\vec{\sigma}  
&   W_0 - (m_N-\sqrt{m_{PS}^2+{\vec p}^2}) W_q  
\end{array}
\right) u^{(s)}
,
\label{eq:dstructure}
\een
where $W_0-W_qi{\slas{q}{}}$ is expressed by a $2\times 2$ block notation;
$\vec{\sigma}$ are the Pauli matrices, and 
${u^{(s)}}^T=(1,0,0,0)$ or $(0,1,0,0)$. 
It is important to observe that the upper components of 
$( W_0 - W_q i{\slas{q}{}}) u^{(s)}$ 
are linear combinations of the 
relevant and irrelevant form factors, 
while the lower components contain only the irrelevant one.
Therefore, we can extract the relevant form factor $W_0$ from the
upper components by subtracting the contribution of 
the irrelevant form factor
$W_q$ with the use of the lower components.

The need for the separation of the contribution of the irrelevant 
form factor was not recognized in the previous studies with the 
direct method\cite{gavela,jlqcd_98}.  The values found in these 
studies correspond to  $W_0-iq_4W_q$ instead of $W_0$.  We examine how 
much this affects the estimate of the matrix elements 
in Sec.~\ref{sec:ndecay}.

Let us add several technical comments: 
(i) The separation procedure described above cannot be applied to
the case of ${\vec p}={\vec k}={\vec 0}$ because of 
vanishing lower components.
(ii) Another possible choice of momenta for disentangling the relevant 
and irrelevant form factors is given by ${\vec k}\ne{\vec 0}$ and 
${\vec p}={\vec 0}$.  In this case, however, we cannot achieve
$-q^2=m_l^2$.

\subsection{Calculational methods}
\label{subsec:calmethod}

The nucleon decay matrix elements 
of eq.~(\ref{eq:ff}) are calculated with two methods, which we refer
to as the direct and the indirect method.  
The former is to extract the matrix elements 
from the three-point
function of the nucleon, the PS meson and the baryon number violating
operator.  
The latter is to estimate them with the aid of chiral lagrangian,
where we have two unknown parameters to be determined
by the lattice QCD calculation. 
 
In the direct method
we calculate the following ratio of the hadron three-point function
divided by the two-point functions:
\ben
R(t,t^\prime)&=&
\frac{\sum_{{\vec x},{\vec x}^\prime} 
      {\rm e}^{i{\vec p}\cdot({\vec x}^\prime-{\vec  x})}
      \la { J}_{PS}({\vec x}^\prime,t^\prime) 
          {\hat {\cal O}}^{\slal{B}{}}_{L,\gamma}({\vec x},t) 
          \bar{ J^\prime}_{N,s}(0) \ra}
     {\sum_{{\vec x},{\vec x}^\prime}
      {\rm e}^{i{\vec p}\cdot({\vec x}^\prime-{\vec x})}
      \la { J}_{PS}({\vec x}^\prime,t^\prime) 
          { J}_{PS}^{\dagger}({\vec x},t) \ra
      \sum_{\vec x}
      \la { J}_{N,s}({\vec x},t) 
      \bar{ J}^\prime_{N,s}(0) \ra }
\sqrt{Z_{PS}} \sqrt{Z_N} \nn \\
&\longrightarrow& 
\frac{1}{L_x L_y L_z}
\la PS(\vec{p})|{\hat {\cal O}}^{\slal{B}{}}_{L,\gamma}
|N^{(s)}({\vec k}={\vec 0})\ra
\;\;\;\;\;\;\;\; t^\prime \gg t \gg 0.
\label{eq:ratio}
\een
Here ${\hat {\cal O}}^{\slal{B}{}}_{L,\gamma}$ denotes
the renormalized operator in the 
naive dimensional regularization(NDR) with the ${\overline{\rm MS}}$ 
subtraction scheme, and 
$\gamma$ and $s$ are spinor indices; we can specify the
spin state of the initial nucleon at rest by choosing
$s=1$ or $2$. $L_xL_yL_z$ is the spatial volume of lattice
in lattice units.
The amplitudes $Z_{PS}$ and $Z_N$ are given by
\ben
  \la PS(\vec{p}) | { J}_{PS}^\dagger({\vec 0},0) | 0 \ra 
  &=& \sqrt{Z_{PS}}, 
\label{eq:z_ps}\\
  \la 0 | { J}_{N,s}({\vec 0},0) | N^{(s^\prime)}({\vec 0}) \ra 
  &=& \sqrt{Z_N} u_s^{(s^\prime)},
\label{eq:z_n}
\een
which can be obtained from the two-point functions
\ben
&&\sum_{{\vec x},{\vec x}^\prime}
      {\rm e}^{i{\vec p}\cdot({\vec x}^\prime-{\vec x})}
      \la { J}_{PS}({\vec x}^\prime,t^\prime) 
          { J}_{PS}^{\dagger}({\vec x},t) \ra, 
\label{eq:z_ps_2pt}\\
&&\sum_{\vec x}
      \la { J}_{N,s}({\vec x},t) 
      \bar{ J}_{N,s}({\vec 0},0) \ra.
\label{eq:z_n_2pt}
\een
We move the baryon number violating operator 
${\hat {\cal O}}^{\slal{B}{}}_{L,\gamma}$ 
in terms of $t$ between the nucleon source placed at $t=0$ and 
the PS meson sink fixed at some $t^\prime$ well separated from 
$t=0$.

We list all the local interpolating fields for the PS meson and the
nucleon required to calculate the independent matrix elements
of eqs.~(\ref{eq:indme_1})$-$(\ref{eq:indme_7}):
\ben
{ J}_{\pi^0}({\vec x},t)&=&
\frac{1}{\sqrt{2}}\left({\bar u}(\vec{x},t)\gamma_5 u(\vec{ x},t) 
           -{\bar d}(\vec{x},t)\gamma_5 d(\vec{ x},t)\right), \\
{ J}_{\pi^+}({\vec x},t)&=&
{\bar d}(\vec{x},t)\gamma_5 u(\vec{ x},t), \\
{ J}_{K^0}({\vec x},t)&=&
{\bar s}(\vec{x},t)\gamma_5 d(\vec{ x},t), \\
{ J}_{K^+}({\vec x},t)&=&
{\bar s}(\vec{x},t)\gamma_5 u(\vec{ x},t), \\
{ J}_{\eta }({\vec x},t)&=&
\frac{1}{\sqrt{6}}\left({\bar u}(\vec{x},t)\gamma_5 u(\vec{ x},t) 
           +{\bar d}(\vec{x},t)\gamma_5 d(\vec{ x},t)
           -2{\bar s}(\vec{x},t)\gamma_5 s(\vec{ x},t)\right), \\
{ J}_{p,s}({\vec x},t)&=&
\epsilon_{ijk}
\left({u^i}^T(\vec{ x},t) C\gamma_5 d^j(\vec{ x},t)\right) u^k_s(\vec{ x},t),\\
{ J}_{n,s}({\vec x},t)&=&
\epsilon_{ijk}
\left({u^i}^T(\vec{ x},t) C\gamma_5 d^j(\vec{ x},t)\right) d^k_s(\vec{ x},t).
\een
We also prepare smeared operators for the nucleon source to overlap
with the lowest energy state dominantly:
\ben
{ J}^\prime_{p,s}(t)&=&
\sum_{{\vec x},{\vec y},{\vec z}}\Psi({\vec x})\Psi({\vec y})\Psi({\vec z})
\epsilon_{ijk}
\left({u^i}^T(\vec{x},t) C\gamma_5 d^j(\vec{y},t)\right) u^k_s(\vec{z},t),
\label{eq:smear_p} \\
{ J}^\prime_{n,s}(t)&=&
\sum_{{\vec x},{\vec y},{\vec z}}\Psi({\vec x})\Psi({\vec y})\Psi({\vec z})
\epsilon_{ijk}
\left({u^i}^T(\vec{x},t) C\gamma_5 d^j(\vec{y},t)\right) d^k_s(\vec{z},t),
\label{eq:smear_n}
\een
where the measured quark wave function in the pion is employed
for the smearing factor $\Psi$,  which is obtained by
\ben
\frac{\sum_{\vec y}\la{\bar d}(\vec{x},t)\gamma_5 u(\vec{0},t) 
          {\bar u}(\vec{y},0)\gamma_5 d(\vec{y},0)\ra}
     {\sum_{\vec y}\la{\bar d}(\vec{0},t)\gamma_5 u(\vec{0},t) 
          {\bar u}(\vec{y},0)\gamma_5 d(\vec{y},0)\ra}
\longrightarrow \Psi({\vec x})
\;\;\;\;\;\;\;\; t\gg 0 
\label{eq:wf_pi}
\een
with configurations fixed to the Coulomb gauge.
Although there is no reason to assume that the wave functions for
the three quarks in the proton is well described by
the quark wave function in the pion, the smeared sources
of eqs.~(\ref{eq:smear_p}) and (\ref{eq:smear_n}) 
work effectively (see Sec.~\ref{sec:ab}).

In the renormalization of the baryon number violating operators 
on the lattice, the explicit chiral symmetry breaking 
in the Wilson quark action causes
mixing between operators with different chiralities.
In eqs.~(\ref{eq:indme_1})$-$(\ref{eq:indme_7}) 
we find two types of operators in terms of chiralities:
\ben
{\cal O}_{RL}=\epsilon_{ijk}{(\psi^i_1}^TCP_{R}\psi^j_2) P_L \psi^{k}_3,
\label{eq:o_rl}\\
{\cal O}_{LL}=\epsilon_{ijk}{(\psi^i_1}^TCP_{L}\psi^j_2) P_L \psi^{k}_3,
\label{eq:o_ll}
\een
where $\psi_{1,2,3}$ represent the quark fields.
Their mixing structures under perturbative renormalization 
up to one-loop level are given by\cite{pt_w}
\ben
{\cal O}_{RL}^{\rm cont}(\mu)&=&
Z(\alpha_s,\mu a){\cal O}_{RL}^{\rm latt}(a)
+\frac{\alpha_s}{4\pi}Z_{mix}{\cal O}_{LL}^{\rm latt}(a)
-\frac{\alpha_s}{4\pi}Z_{mix}^\prime{\cal O}_{\gamma_\mu L}^{\rm latt}(a),
\label{eq:z_rl}\\
{\cal O}_{LL}^{\rm cont}(\mu)&=&
Z(\alpha_s,\mu a){\cal O}_{LL}^{\rm latt}(a)
+\frac{\alpha_s}{4\pi}Z_{mix}{\cal O}_{RL}^{\rm latt}(a)
+\frac{\alpha_s}{4\pi}Z_{mix}^\prime{\cal O}_{\gamma_\mu L}^{\rm latt}(a), 
\label{eq:z_ll}
\een
where the overall factor $Z(\alpha_s, \mu a)$ has the form
\be
Z(\alpha_s,\mu a) = 1+\frac{\alpha_s}{4\pi}
\left(4{\rm ln}(\mu a)+\Delta_{\slal{B}{}}\right),
\label{eq:z_overall} \\
\ee
with $\mu$ the renormalization scale, and the additional operator
${\cal O}_{\gamma_\mu L}$ is defined by
\be
{\cal O}_{\gamma_\mu L}=\epsilon_{ijk}
{(\psi^i_1}^TC\gamma_\mu\gamma_5 \psi^j_2) P_L \gamma_\mu \psi^{k}_3.
\ee
Employing the $\overline {\rm MS}$ subtraction scheme 
with the naive dimensional regularization for the continuum theory,
we have reevaluated the finite constants and found
\ben
\Delta_{\slal{B}{}}&=&-34.11\hspace{24pt}{\rm for\hspace{12pt}NDR},  \\
Z_{mix}&=&3.21, 
\label{eq:z_mix}\\
Z_{mix}^\prime&=&-0.803,
\label{eq:z_mix_prime}
\een
where the errors are $\pm 1$ in the last digit.
The value of $\Delta_{\slal{B}{}}$ depends on the renormalization
scheme in the continuum,
while $Z_{mix}$ and $Z^\prime_{mix}$ are independent. We present the 
integral for $\Delta_{\slal{B}{}}$ in the dimensional reduction(DRED) 
scheme and those for $Z_{mix}$ and $Z^\prime_{mix}$ 
in Appendix, where we give a detailed description
for the one-loop perturbative calculation of the renormalization factors.
With the use of the KLM normalization of quark fields\cite{klm} 
and the tadpole improvement\cite{tadimp},
the overall renormalization factor of eq.~(\ref{eq:z_overall})
is rewritten as
\be
Z(\alpha_s,\mu a)=\left(1-\frac{3K}{4K_c}\right)^{\frac{3}{2}}
\left[1+\frac{\alpha_s}{4\pi}
\left(4{\rm ln}(\mu a)+\Delta_{\slal{B}{}}
+\frac{3}{2}\pi\times 5.457 \right)\right].
\label{eq:z_overall_tad}
\ee
Here $K_c$ is the critical hopping parameter at which the
pion mass vanishes. We use
\be
\frac{1}{8K_c}=1-5.457\alpha_s/4
\ee 
in Ref.~\cite{pt_kc} for the perturbative estimate of $K_c$.

Let us turn to the indirect method.
The baryon number violating operators constructed
in chiral lagrangian 
contains  two unknown coefficients $\alpha$ and $\beta$ defined by
\ben
\la 0 | \epsilon_{ijk}({u^i}^T C P_R d^j) P_L u^k 
| p^{(s)} \ra &=& \alpha P_L u^{(s)}, 
\label{eq:alpha}\\
\la 0 | \epsilon_{ijk}({u^i}^T C P_L d^j) P_L u^k 
| p^{(s)} \ra &=& \beta  P_L u^{(s)},
\label{eq:beta}
\een
where operators are renormalized in the NDR scheme
with  the use of the renormalization factors
of eqs.~(\ref{eq:z_mix})$-$ (\ref{eq:z_overall_tad}).
These matrix elements are obtained from the two-point
functions:
\ben
R^{\alpha\beta}(t)&=&
\frac{\sum_{\vec x}
      \la \epsilon_{ijk}({u^i}^T C P_{R,L} d^j) P_L u^k({\vec x},t)
      \bar{ J}^\prime_{p,s}(0)\ra}
     {\sum_{\vec x}
      \la { J}_{p,s}({\vec x},t)
      \bar{ J}^\prime_{p,s}(0)\ra}
      \sqrt{Z_N} \nn\\
      &\longrightarrow&
      \la 0 | \epsilon_{ijk}(u^i C P_{R,L} d^j) P_L u^k | p^{(s)} \ra
\;\;\;\;\;\;\;\; t\gg 0 . 
\label{eq:ratio_ab}
\een
Incorporating the $\alpha$ and $\beta$ values
determined by the lattice calculation  in the
tree-level results of chiral lagrangian, 
we can evaluates the
nucleon decay matrix elements of eqs.~(\ref{eq:indme_1})$-$(\ref{eq:indme_7}).

\section{Tree-level results of chiral lagrangian}
\label{sec:chpt}

For some nucleon decay matrix elements, 
tree-level results of chiral lagrangian have already been 
given in Refs.~\cite{chpt_gut,chpt_susy}, 
which are obtained with the use of the on-shell
condition of the out-going leptons: $-q^2=m_l^2$ and $i\slas{q}{}v_l=m_l v_l$. 
In our lattice calculations, however, the lepton momentum is 
generally off the mass shell.  Hence we need to understand the 
$q$ dependence for an extrapolation of the matrix elements to the 
physical point. 
In this section we present the tree-level results for all the
independent matrix elements in 
eqs.~(\ref{eq:indme_1})$-$(\ref{eq:indme_7}) with the explicit expressions
of $q$ dependences.

We first define the chiral lagrangian for baryon-meson strong
interactions following the notation of Ref.~\cite{chpt_gut}. 
The PS meson and baryon fields are given by 
\ben
\phi&=&\left(\begin{array}{ccc}
     \frac{1}{\sqrt{2}}\pi^0+\frac{1}{\sqrt{6}}\eta &\pi^+&K^+ \\
     \pi^-&-\frac{1}{\sqrt{2}}\pi^0+\frac{1}{\sqrt{6}}\eta &K^0 \\
     K^-&{\bar K}^0&-\frac{2}{\sqrt{6}}\eta \\
             \end{array} \right), \\
B&=&\left(\begin{array}{ccc}
     \frac{1}{\sqrt{2}}\Sigma^0+\frac{1}{\sqrt{6}}\Lambda^0&\Sigma^+&p \\
     \Sigma^-&-\frac{1}{\sqrt{2}}\Sigma^0+\frac{1}{\sqrt{6}}\Lambda^0&n \\
     \Xi^-&\Xi^0&-\frac{2}{\sqrt{6}}\Lambda^0 \\
          \end{array} \right).
\een
In terms of $\phi$ we define the $3\times3$ special unitary matrices:
\ben
\Sigma={\rm exp}\left(\frac{2i\phi}{f}\right),\\
\xi={\rm exp}\left(\frac{i\phi}{f}\right),
\een
where $f$ is the pion decay constant.
Under SU(3)$_L\times$SU(3)$_R$ the meson and baryon fields
transform as 
\ben
\Sigma\rightarrow L\Sigma R^\dagger,\\
B\rightarrow UBU^\dagger,
\een
where $L$ is an element of SU(3)$_L$ and $R$ is an element of 
SU(3)$_R$; $U$ is defined through the transformation properties 
of $\xi$:
\be
\xi\rightarrow L\xi U^\dagger=U\xi R^\dagger.
\ee

The lowest order of the SU(3)$_L\times$SU(3)$_R$ invariant
chiral lagrangian is given by
\ben
{\cal L}_0&=&\frac{f^2}{8}\tr(\partial_\mu\Sigma)(\partial_\mu\Sigma^\dagger)
             +\tr{\bar B}(\gamma_\mu\partial_\mu+M_B)B\nn\\
          && +\frac{1}{2}\tr{\bar B}\gamma_\mu[\xi\partial_\mu\xi^\dagger
             +\xi^\dagger\partial_\mu\xi]B 
             +\frac{1}{2}\tr{\bar B}\gamma_\mu B[(\partial_\mu\xi)\xi^\dagger
             +(\partial_\mu\xi^\dagger)\xi] \nn \\
          && -\frac{1}{2}(D-F)\tr{\bar B}\gamma_\mu\gamma_5 B
             [(\partial_\mu\xi)\xi^\dagger
             -(\partial_\mu\xi^\dagger)\xi] \nn \\
          && +\frac{1}{2}(D+F)\tr{\bar B}\gamma_\mu\gamma_5 
             [\xi\partial_\mu\xi^\dagger
             -\xi^\dagger\partial_\mu\xi]B
\een
on the Euclidean space-time. 
Quark mass contributions can be
included by adding the symmetry-breaking term 
\ben
{\cal L}_1&=&-v^3\tr(\Sigma^\dagger M_q+M_q\Sigma) \nn \\
          && -a_1\tr{\bar B}(\xi^\dagger M_q\xi^\dagger
             +\xi M_q\xi)B 
             -a_2\tr{\bar B}B(\xi^\dagger M_q\xi^\dagger
             +\xi M_q\xi) \nn\\
          && -b_1\tr{\bar B}\gamma_5(\xi^\dagger M_q\xi^\dagger
             -\xi M_q\xi)B 
             -b_2\tr{\bar B}\gamma_5B(\xi^\dagger M_q\xi^\dagger,
             -\xi M_q\xi), 
\een
where
\be
M_q=\left(\begin{array}{ccc}
       m_u & 0 & 0 \\
       0 & m_d & 0 \\
       0 & 0 & m_s \\
          \end{array} \right).
\ee
The parameter $v$ is related to the meson mass by
\be
v= \frac{f^2 m_{\pi^{\pm,0}}^2}{4(m_u+m_d)}
 = \frac{f^2 m_{K^{\pm}}^2}{4(m_u+m_s)}
 = \frac{f^2 m_{K^{0}}^2}{4(m_d+m_s)}
 = \frac{3f^2 m_{\eta }^2}{4(m_u+m_d+4m_s)}.
\ee
Experimental results for the semileptonic baryon decays 
give $F=0.47$ and $D=0.80$\cite{fd}.
The symmetry-breaking parameters $a_1$ and $a_2$ are
estimated from mass splittings among the octet baryons;
$b_1$ and $b_2$, on the other hand, are not well determined since
they do not contribute to the baryon masses. 
The parameters $v$, $a_1$ and $a_2$ have no contribution to the
tree-level results for the nucleon decay matrix elements.

Let us consider the construction of the operators of 
eqs.~(\ref{eq:op_ns_1})$-$(\ref{eq:op_s_6}), which are
written in the quark fields, with the 
meson and baryon fields. 
The operators transform
under SU(3)$_L\times$SU(3)$_R$ as
\ben
(3,{\bar 3}) &:& {\cal O}_d^{(1)},{\tilde {\cal O}}_d^{(1)}
                ,{\tilde {\cal O}}_d^{(5)}, \\
({\bar 3},3) &:& {\cal O}_d^{(2)},{\tilde {\cal O}}_d^{(2)}, \\
(8,1)        &:& {\cal O}_d^{(3)},{\tilde {\cal O}}_d^{(3)}
                ,{\tilde {\cal O}}_d^{(6)}, \\
(1,8)        &:& {\cal O}_d^{(4)},{\tilde {\cal O}}_d^{(4)}. 
\een
These transformation properties are realized by
$\xi B\xi\in (3,{\bar 3})$,    
$\xi^\dagger B\xi^\dagger\in ({\bar 3},3)$
$\xi B\xi^\dagger\in (8,1)$ and $\xi^\dagger B\xi\in (1,8)$,
with which we can express the operators of 
eqs.~(\ref{eq:op_ns_1})$-$(\ref{eq:op_s_6}) as
\ben
{\cal O}^{(1)}_{d}&=&\alpha
\left({\bar e}^c_{dL}\tr{\cal F}\xi B_L\xi
-{\bar \nu}^c_{dL}\tr{\cal F}^\prime\xi B_L\xi\right),
\label{eq:op_ns_1_cl}\\
{\cal O}^{(2)}_{d}&=&\alpha
{\bar e}^c_{dR}\tr{\cal F}\xi^\dagger B_R\xi^\dagger,
\label{eq:op_ns_2_cl}\\
{\cal O}^{(3)}_{d}&=&\beta
\left({\bar e}^c_{dL}\tr{\cal F}\xi B_L\xi^\dagger
-{\bar \nu}^c_{dL}\tr{\cal F}^\prime\xi B_L\xi^\dagger\right),
\label{eq:op_ns_3_cl}\\
{\cal O}^{(4)}_{d}&=&\beta
{\bar e}^c_{dR}\tr{\cal F}\xi^\dagger B_R\xi,
\label{eq:op_ns_4_cl}\\
{\tilde {\cal O}}^{(1)}_{d}&=&\alpha
\left({\bar e}^c_{dL}\tr{\tilde {\cal F}}\xi B_L\xi
-{\bar \nu}^c_{dL}\tr{\tilde {\cal F}}^\prime\xi B_L\xi\right),
\label{eq:op_s_1_cl}\\
{\tilde {\cal O}}^{(2)}_{d}&=&\alpha
{\bar e}^c_{dR}\tr{\tilde {\cal F}}\xi^\dagger B_R\xi^\dagger,
\label{eq:op_s_2_cl}\\
{\tilde {\cal O}}^{(3)}_{d}&=&\beta
\left({\bar e}^c_{dL}\tr{\tilde {\cal F}}\xi B_L\xi^\dagger
-{\bar \nu}^c_{dL}\tr{\tilde {\cal F}}^\prime\xi B_L\xi^\dagger\right),
\label{eq:op_s_3_cl}\\
{\tilde {\cal O}}^{(4)}_{d}&=&\beta
{\bar e}^c_{dR}\tr{\tilde {\cal F}}\xi^\dagger B_R\xi,
\label{eq:op_s_4_cl}\\
{\tilde {\cal O}}^{(5)}_{d}&=&\alpha
{\bar \nu}^c_{dL}\tr{\tilde {\cal F}}^{\prime\prime}\xi B_L\xi,
\label{eq:op_s_5_cl}\\
{\tilde {\cal O}}^{(6)}_{d}&=&\beta
{\bar \nu}^c_{dL}\tr{\tilde {\cal F}}^{\prime\prime}\xi B_L\xi^\dagger,
\label{eq:op_s_6_cl}
\een
where $\alpha$ and $\beta$, which are already defined 
in eqs.~(\ref{eq:alpha}) and (\ref{eq:beta}),
are unknown
coefficients associated with the $(3,{\bar 3})$ and $({\bar 3},3)$
operators and the $(8,1)$ and $(1,8)$ operators respectively; 
${\cal F}$, ${\cal F}^\prime$, ${\tilde {\cal F}}$, 
${\tilde {\cal F}}^\prime$ and ${\tilde {\cal F}}^{\prime\prime}$
are projection matrices in the flavor space,
\ben
&&{\cal F}=\left(\begin{array}{ccc}
          0 & 0 & 0 \\
          0 & 0 & 0 \\
          1 & 0 & 0 \\
          \end{array} \right), 
{\cal F}^\prime=\left(\begin{array}{ccc}
          0 & 0 & 0 \\
          0 & 0 & 0 \\
          0 & 1 & 0 \\
          \end{array} \right), \\
&&{\tilde {\cal F}}=\left(\begin{array}{ccc}
          0 & 0 & 0 \\
         -1 & 0 & 0 \\
          0 & 0 & 0 \\
          \end{array} \right), 
{\tilde {\cal F}}^\prime=\left(\begin{array}{ccc}
          0 & 0 & 0 \\
          0 &-1 & 0 \\
          0 & 0 & 0 \\
          \end{array} \right), 
{\tilde {\cal F}}^{\prime\prime}=\left(\begin{array}{ccc}
          0 & 0 & 0 \\
          0 & 0 & 0 \\
          0 & 0 & 1 \\
          \end{array} \right).
\een

We can now apply the chiral lagrangian ${\cal L}_{0}+{\cal L}_{1}$ 
and the baryon number violating operators of 
eqs.~(\ref{eq:op_ns_1_cl})$-$(\ref{eq:op_s_6_cl})
to calculating the nucleon decay matrix elements.
Expanding the lagrangian and the operators in terms of
the meson and baryon fields, we obtain the following tree-level results
for the independent matrix elements of 
eqs.~(\ref{eq:indme_1})$-$(\ref{eq:indme_7}):
\ben
\langle \pi^0|(ud_R) u_L|p\rangle
&=&\alpha P_L u_p\left[\frac{1}{\sqrt{2}f}
-\frac{D+F}{\sqrt{2}f}\frac{-q^2+m_N^2}{-q^2-m_N^2} 
-\frac{4b_1}{\sqrt{2}f}\frac{m_u m_N}{-q^2-m_N^2}\right] \nn\\
&&-\alpha P_L i\slas{q}{}u_p\left[
\frac{D+F}{\sqrt{2}f}\frac{2m_N}{-q^2-m_N^2} 
+\frac{4b_1}{\sqrt{2}f}\frac{m_u}{-q^2-m_N^2}\right],
\label{eq:chpt_1_rl}\\ 
\langle \pi^0|(ud_L) u_L|p\rangle
&=&\beta P_L u_p\left[\frac{1}{\sqrt{2}f}
-\frac{D+F}{\sqrt{2}f}\frac{-q^2+m_N^2}{-q^2-m_N^2} 
-\frac{4b_1}{\sqrt{2}f}\frac{m_u m_N}{-q^2-m_N^2}\right] \nn\\
&&-\beta P_L i\slas{q}{}u_p\left[
\frac{D+F}{\sqrt{2}f}\frac{2m_N}{-q^2-m_N^2} 
+\frac{4b_1}{\sqrt{2}f}\frac{m_u}{-q^2-m_N^2}\right],
\label{eq:chpt_1_ll}\\ 
%
\langle \pi^+|(ud_R) d_L|p\rangle
&=&\alpha P_L u_p\left[\frac{1}{f}
-\frac{D+F}{f}\frac{-q^2+m_N^2}{-q^2-m_N^2} 
-\frac{2b_1}{f}\frac{(m_u+m_d) m_N}{-q^2-m_N^2}\right] \nn\\
&&-\alpha P_L i\slas{q}{}u_p\left[
\frac{D+F}{f}\frac{2m_N}{-q^2-m_N^2} 
+\frac{2b_1}{f}\frac{m_u+m_d}{-q^2-m_N^2}\right],
\label{eq:chpt_2_rl}\\
\langle \pi^+|(ud_L) d_L|p\rangle
&=&\beta P_L u_p\left[\frac{1}{f}
-\frac{D+F}{f}\frac{-q^2+m_N^2}{-q^2-m_N^2} 
-\frac{2b_1}{f}\frac{(m_u+m_d) m_N}{-q^2-m_N^2}\right] \nn\\
&&-\beta P_L i\slas{q}{}u_p\left[
\frac{D+F}{f}\frac{2m_N}{-q^2-m_N^2} 
+\frac{2b_1}{f}\frac{m_u+m_d}{-q^2-m_N^2}\right],
\label{eq:chpt_2_ll}\\
%
\langle K^0|(us_R) u_L|p\rangle 
&=&\alpha P_L u_p\left[-\frac{1}{f}
+\frac{D-F}{f}\frac{-q^2+m_N m_\Sigma}{-q^2-m_\Sigma^2} 
+\frac{2b_2}{f}\frac{(m_d+m_s) m_\Sigma}{-q^2-m_\Sigma^2}\right] \nn\\
&&-\alpha P_L i\slas{q}{}u_p\left[
-\frac{D-F}{f}\frac{m_N+m_\Sigma}{-q^2-m_\Sigma^2} 
-\frac{2b_2}{f}\frac{m_d+m_s}{-q^2-m_\Sigma^2}\right],
\label{eq:chpt_3_rl}\\
\langle K^0|(us_L) u_L|p\rangle 
&=&\beta P_L u_p\left[\frac{1}{f}
+\frac{D-F}{f}\frac{-q^2+m_N m_\Sigma}{-q^2-m_\Sigma^2} 
+\frac{2b_2}{f}\frac{(m_d+m_s) m_\Sigma}{-q^2-m_\Sigma^2}\right] \nn\\
&&-\beta P_L i\slas{q}{}u_p\left[
-\frac{D-F}{f}\frac{m_N+m_\Sigma}{-q^2-m_\Sigma^2} 
-\frac{2b_2}{f}\frac{m_d+m_s}{-q^2-m_\Sigma^2}\right],
\label{eq:chpt_3_ll}\\
%
\langle K^+|(us_R) d_L|p\rangle
&=&\alpha P_L u_p\left[
-\frac{D-F}{2f}\frac{-q^2+m_N m_\Sigma}{-q^2-m_\Sigma^2}
-\frac{D+3F}{6f}\frac{-q^2+m_N m_\Lambda}{-q^2-m_\Lambda^2} \right.\nn\\
&&\left.
-\frac{b_2}{f}\frac{(m_u+m_s) m_\Sigma}{-q^2-m_\Sigma^2} 
+\frac{b_2-2b_1}{3f}\frac{(m_u+m_s) m_\Lambda}{-q^2-m_\Lambda^2}\right] \nn\\
&&-\alpha P_L i\slas{q}{}u_p\left[
+\frac{D-F}{2f}\frac{m_N+m_\Sigma}{-q^2-m_\Sigma^2}
+\frac{D+3F}{6f}\frac{m_N+m_\Lambda}{-q^2-m_\Lambda^2} \right.\nn\\
&&\left.
+\frac{b_2}{f}\frac{m_u+m_s}{-q^2-m_\Sigma^2} 
-\frac{b_2-2b_1}{3f}\frac{m_u+m_s}{-q^2-m_\Lambda^2}\right],
\label{eq:chpt_4_rl}\\
\langle K^+|(us_L) d_L|p\rangle
&=&\beta P_L u_p\left[
-\frac{D-F}{2f}\frac{-q^2+m_N m_\Sigma}{-q^2-m_\Sigma^2}
-\frac{D+3F}{6f}\frac{-q^2+m_N m_\Lambda}{-q^2-m_\Lambda^2} \right.\nn\\
&&\left.
-\frac{b_2}{f}\frac{(m_u+m_s) m_\Sigma}{-q^2-m_\Sigma^2} 
+\frac{b_2-2b_1}{3f}\frac{(m_u+m_s) m_\Lambda}{-q^2-m_\Lambda^2}\right] \nn\\
&&-\beta P_L i\slas{q}{}u_p\left[
+\frac{D-F}{2f}\frac{m_N+m_\Sigma}{-q^2-m_\Sigma^2}
+\frac{D+3F}{6f}\frac{m_N+m_\Lambda}{-q^2-m_\Lambda^2} \right.\nn\\
&&\left.
+\frac{b_2}{f}\frac{m_u+m_s}{-q^2-m_\Sigma^2} 
-\frac{b_2-2b_1}{3f}\frac{m_u+m_s}{-q^2-m_\Lambda^2}\right], 
\label{eq:chpt_4_ll}\\
%
\langle K^+|(ud_R) s_L|p\rangle 
&=&\alpha P_L u_p\left[\frac{1}{f}
-\frac{D+3F}{3f}\frac{-q^2+m_N m_\Lambda}{-q^2-m_\Lambda^2} \right.\nn\\
&&\left.
+\frac{2(b_2-2b_1)}{3f}\frac{(m_u+m_s) m_\Lambda}
{-q^2-m_\Lambda^2}\right] \nn\\
&&-\alpha P_L i\slas{q}{}u_p\left[
+\frac{D+3F}{3f}\frac{m_N+m_\Lambda}{-q^2-m_\Lambda^2} \right.\nn\\
&&\left.
-\frac{2(b_2-2b_1)}{3f}\frac{m_u+m_s}{-q^2-m_\Lambda^2}\right],
\label{eq:chpt_5_rl}\\
\langle K^+|(ud_L) s_L|p\rangle 
&=&\beta P_L u_p\left[\frac{1}{f}
-\frac{D+3F}{3f}\frac{-q^2+m_N m_\Lambda}{-q^2-m_\Lambda^2} \right.\nn\\
&&\left.
+\frac{2(b_2-2b_1)}{3f}\frac{(m_u+m_s) m_\Lambda}
{-q^2-m_\Lambda^2}\right] \nn\\
&&-\beta P_L i\slas{q}{}u_p\left[
+\frac{D+3F}{3f}\frac{m_N+m_\Lambda}{-q^2-m_\Lambda^2} \right.\nn\\
&&\left.
-\frac{2(b_2-2b_1)}{3f}\frac{m_u+m_s}{-q^2-m_\Lambda^2}\right],
\label{eq:chpt_5_ll}\\
%
\langle K^0|(us_R) d_L|n\rangle
&=&\alpha P_L u_p\left[-\frac{1}{f}
+\frac{D-F}{2f}\frac{-q^2+m_N m_\Sigma}{-q^2-m_\Sigma^2}
-\frac{D+3F}{6f}\frac{-q^2+m_N m_\Lambda}{-q^2-m_\Lambda^2} \right.\nn\\
&&\left.
+\frac{b_2}{f}\frac{(m_d+m_s) m_\Sigma}{-q^2-m_\Sigma^2} 
+\frac{b_2-2b_1}{3f}\frac{(m_d+m_s) m_\Lambda}{-q^2-m_\Lambda^2}\right] \nn\\
&&-\alpha P_L i\slas{q}{}u_p\left[
-\frac{D-F}{2f}\frac{m_N+m_\Sigma}{-q^2-m_\Sigma^2}
+\frac{D+3F}{6f}\frac{m_N+m_\Lambda}{-q^2-m_\Lambda^2} \right.\nn\\
&&\left.
-\frac{b_2}{f}\frac{m_d+m_s}{-q^2-m_\Sigma^2} 
-\frac{b_2-2b_1}{3f}\frac{m_d+m_s}{-q^2-m_\Lambda^2}\right], 
\label{eq:chpt_6_rl}\\
\langle K^0|(us_L) d_L|n\rangle
&=&\beta P_L u_p\left[\frac{1}{f}
+\frac{D-F}{2f}\frac{-q^2+m_N m_\Sigma}{-q^2-m_\Sigma^2}
-\frac{D+3F}{6f}\frac{-q^2+m_N m_\Lambda}{-q^2-m_\Lambda^2} \right.\nn\\
&&\left.
+\frac{b_2}{f}\frac{(m_d+m_s) m_\Sigma}{-q^2-m_\Sigma^2} 
+\frac{b_2-2b_1}{3f}\frac{(m_d+m_s) m_\Lambda}{-q^2-m_\Lambda^2}\right] \nn\\
&&-\beta P_L i\slas{q}{}u_p\left[
-\frac{D-F}{2f}\frac{m_N+m_\Sigma}{-q^2-m_\Sigma^2}
+\frac{D+3F}{6f}\frac{m_N+m_\Lambda}{-q^2-m_\Lambda^2} \right.\nn\\
&&\left.
-\frac{b_2}{f}\frac{m_d+m_s}{-q^2-m_\Sigma^2} 
-\frac{b_2-2b_1}{3f}\frac{m_d+m_s}{-q^2-m_\Lambda^2}\right], 
\label{eq:chpt_6_ll}\\
%
\langle \eta |(ud_R) u_L|p\rangle
&=&\alpha P_L u_p\left[-\frac{1}{\sqrt{6}f}
+\frac{D-3F}{\sqrt{6}f}\frac{-q^2+m_N^2}{-q^2-m_N^2} \right.\nn\\
&&\left.
-\frac{4(b_1 m_u-2 b_2 m_s)}{\sqrt{6}f}\frac{m_N}{-q^2-m_N^2}\right] \nn\\
&&-\alpha P_L i\slas{q}{}u_p\left[
-\frac{D-3F}{\sqrt{6}f}\frac{2m_N}{-q^2-m_N^2} \right.\nn\\
&&\left.
+\frac{4(b_1 m_u-2 b_2 m_s)}{\sqrt{6}f}\frac{1}{-q^2-m_N^2}\right],
\label{eq:chpt_7_rl}\\ 
\langle \eta |(ud_L) u_L|p\rangle
&=&\beta P_L u_p\left[\frac{3}{\sqrt{6}f}
+\frac{D-3F}{\sqrt{6}f}\frac{-q^2+m_N^2}{-q^2-m_N^2} \right.\nn\\
&&\left.
-\frac{4(b_1 m_u-2 b_2 m_s)}{\sqrt{6}f}\frac{m_N}{-q^2-m_N^2}\right] \nn\\
&&-\beta P_L i\slas{q}{}u_p\left[
-\frac{D-3F}{\sqrt{6}f}\frac{2m_N}{-q^2-m_N^2} \right.\nn\\
&&\left.
+\frac{4(b_1 m_u-2 b_2 m_s)}{\sqrt{6}f}\frac{1}{-q^2-m_N^2}\right],
\label{eq:chpt_7_ll}
\een
where we use $\langle PS |(\psi_1 {\psi_2}_{R,L}) {\psi_3}_L|N\rangle$ 
as a shortened form of $\langle PS|\epsilon_{ijk}
({\psi_1^i}^T CP_{R,L}{\psi_2^j}) 
P_L{\psi_3^k}|N\rangle$; $q$ dependences of the matrix elements
are retained without applying
the on-shell condition of the out-going leptons 
$-q^2=m_l^2$, $i\slas{q}{}v_l=m_l v_l$.
These expressions are considerably simplified 
if we employ the approximations of
$m_\Sigma\simeq m_\Lambda \equiv m_B$, $m_{u,d}\ll m_s$, 
$m_s/m_B\ll m_N/m_B$, $b_{1,2}\sim O(1)$ and $-q^2\ll m_{N,B}$:
\ben
\langle \pi^0|(ud_R) u_L|p\rangle
&\simeq&
\alpha P_L u_p\left[\frac{1}{\sqrt{2}f}
+\frac{D+F}{\sqrt{2}f}
\left(1+2\frac{(-q^2)}{m_N^2}+2\frac{(-q^2)^2}{m_N^4}\right)\right]+O(q^6),
\label{eq:chpt_1_rl_q}\\ 
\langle \pi^0|(ud_L) u_L|p\rangle
&\simeq&
\beta P_L u_p\left[\frac{1}{\sqrt{2}f}
+\frac{D+F}{\sqrt{2}f}
\left(1+2\frac{(-q^2)}{m_N^2}+2\frac{(-q^2)^2}{m_N^4}\right)\right]+O(q^6),
\label{eq:chpt_1_ll_q}\\ 
%
\langle \pi^+|(ud_R) d_L|p\rangle
&\simeq&
\alpha P_L u_p\left[\frac{1}{f}
+\frac{D+F}{f}
\left(1+2\frac{(-q^2)}{m_N^2}+2\frac{(-q^2)^2}{m_N^4}\right)\right]+O(q^6),
\label{eq:chpt_2_rl_q}\\
\langle \pi^+|(ud_L) d_L|p\rangle
&\simeq&
\beta P_L u_p\left[\frac{1}{f}
+\frac{D+F}{f}
\left(1+2\frac{(-q^2)}{m_N^2}+2\frac{(-q^2)^2}{m_N^4}\right)\right]+O(q^6),
\label{eq:chpt_2_ll_q}\\
%
\langle K^0|(us_R) u_L|p\rangle 
&\simeq&
\alpha P_L u_p\left[-\frac{1}{f}
-\frac{D-F}{f}
\left\{\frac{m_N}{m_B} \right.\right.\nn\\
&&\left.\left.+\frac{m_N+m_B}{m_B}
\left(\frac{(-q^2)}{m_B^2}+\frac{(-q^2)^2}{m_B^4}\right)\right\}\right]+O(q^6),
\label{eq:chpt_3_rl_q}\\
\langle K^0|(us_L) u_L|p\rangle 
&\simeq&
\beta P_L u_p\left[\frac{1}{f}
-\frac{D-F}{f}
\left\{\frac{m_N}{m_B} \right.\right.\nn\\
&&\left.\left.+\frac{m_N+m_B}{m_B}
\left(\frac{(-q^2)}{m_B^2}+\frac{(-q^2)^2}{m_B^4}\right)\right\}\right]+O(q^6),
\label{eq:chpt_3_ll_q}\\
%
\langle K^+|(us_R) d_L|p\rangle
&\simeq&
\alpha P_L u_p\left[
+\frac{2D}{3f}
\left\{\frac{m_N}{m_B}+\frac{m_N+m_B}{m_B}
\left(\frac{(-q^2)}{m_B^2}+\frac{(-q^2)^2}{m_B^4}\right)\right\}\right]+O(q^6),
\label{eq:chpt_4_rl_q}\\
\langle K^+|(us_L) d_L|p\rangle
&\simeq&
\beta P_L u_p\left[
+\frac{2D}{3f}
\left\{\frac{m_N}{m_B}+\frac{m_N+m_B}{m_B}
\left(\frac{(-q^2)}{m_B^2}+\frac{(-q^2)^2}{m_B^4}\right)\right\}\right]+O(q^6),
\label{eq:chpt_4_ll_q}\\
%
\langle K^+|(ud_R) s_L|p\rangle 
&\simeq&
\alpha P_L u_p\left[\frac{1}{f}
+\frac{D+3F}{3f}
\left\{\frac{m_N}{m_B} \right.\right.\nn\\
&&\left.\left.+\frac{m_N+m_B}{m_B}
\left(\frac{(-q^2)}{m_B^2}+\frac{(-q^2)^2}{m_B^4}\right)\right\}\right]+O(q^6),
\label{eq:chpt_5_rl_q}\\
\langle K^+|(ud_L) s_L|p\rangle 
&\simeq&
\beta P_L u_p\left[\frac{1}{f}
+\frac{D+3F}{3f}
\left\{\frac{m_N}{m_B} \right.\right.\nn\\
&&\left.\left.+\frac{m_N+m_B}{m_B}
\left(\frac{(-q^2)}{m_B^2}+\frac{(-q^2)^2}{m_B^4}\right)\right\}\right]+O(q^6),
\label{eq:chpt_5_ll_q}\\
%
\langle K^0|(us_R) d_L|n\rangle
&\simeq&
\alpha P_L u_p\left[-\frac{1}{f}
-\frac{D-3F}{3f}
\left\{\frac{m_N}{m_B} \right.\right.\nn\\
&&\left.\left.+\frac{m_N+m_B}{m_B}
\left(\frac{(-q^2)}{m_B^2}+\frac{(-q^2)^2}{m_B^4}\right)\right\}\right]+O(q^6),
\label{eq:chpt_6_rl_q}\\
\langle K^0|(us_L) d_L|n\rangle
&\simeq&
\beta P_L u_p\left[\frac{1}{f}
-\frac{D-3F}{3f}
\left\{\frac{m_N}{m_B} \right.\right.\nn\\
&&\left.\left.+\frac{m_N+m_B}{m_B}
\left(\frac{(-q^2)}{m_B^2}+\frac{(-q^2)^2}{m_B^4}\right)\right\}\right]+O(q^6),
\label{eq:chpt_6_ll_q}\\
%
\langle \eta |(ud_R) u_L|p\rangle
&\simeq&
\alpha P_L u_p\left[-\frac{1}{\sqrt{6}f}
-\frac{D-3F}{\sqrt{6}f}
\left(1+2\frac{(-q^2)}{m_N^2}+2\frac{(-q^2)^2}{m_N^4}\right)\right]+O(q^6),
\label{eq:chpt_7_rl_q}\\ 
\langle \eta |(ud_L) u_L|p\rangle
&\simeq&
\beta P_L u_p\left[\frac{3}{\sqrt{6}f}
-\frac{D-3F}{\sqrt{6}f}
\left(1+2\frac{(-q^2)}{m_N^2}+2\frac{(-q^2)^2}{m_N^4}\right)\right]+O(q^6),
\label{eq:chpt_7_ll_q}
\een
where we present only the relevant terms.

\section{Details of numerical simulation} 
\label{sec:parameter}

\subsection{Data sets}

Our calculation is carried out with the Wilson quark action in quenched QCD 
at $\beta=6.0$ on a $28^2\times 48\times 80$ lattice. 
Gauge configurations are generated with the single plaquette
action separated by 2000 pseudo heat-bath sweeps. 
We employ 20 configurations for the measurement of the
quark wave function in the pion, which is used for the nucleon 
smeared source, 
after the thermalization of 22000 sweeps, and then analyzed
the next 100 configurations for the calculation of the 
nucleon decay matrix elements.
The four hopping parameters $K=0.15620$, $0.15568$, 
$0.15516$ and $0.15464$ 
are adopted such that the physical point for the $K$ meson can be
interpolated.
The critical hopping parameter $K_c=0.15714(1)$ is determined by
extrapolating the results of $m_\pi^2$ at the four hopping parameters 
linearly in $1/2K$ to $m_\pi^2=0$. The $\rho$ meson mass at the 
chiral limit is used to determine the inverse lattice spacing 
$a^{-1}=2.30(4)$GeV with $m_\rho=770$MeV as input.
The strange quark mass $m_s a=0.0464(16)$($K_s=0.15488(7)$), 
which is estimated from the
experimental ratio $m_K/m_\rho=0.644$, is in the middle of $K=0.15516$ and
$K=0.15464$.

\subsection{Calculational procedure}

\begin{figure}[h]
\centering{
\hskip -0.0cm
\psfig{file=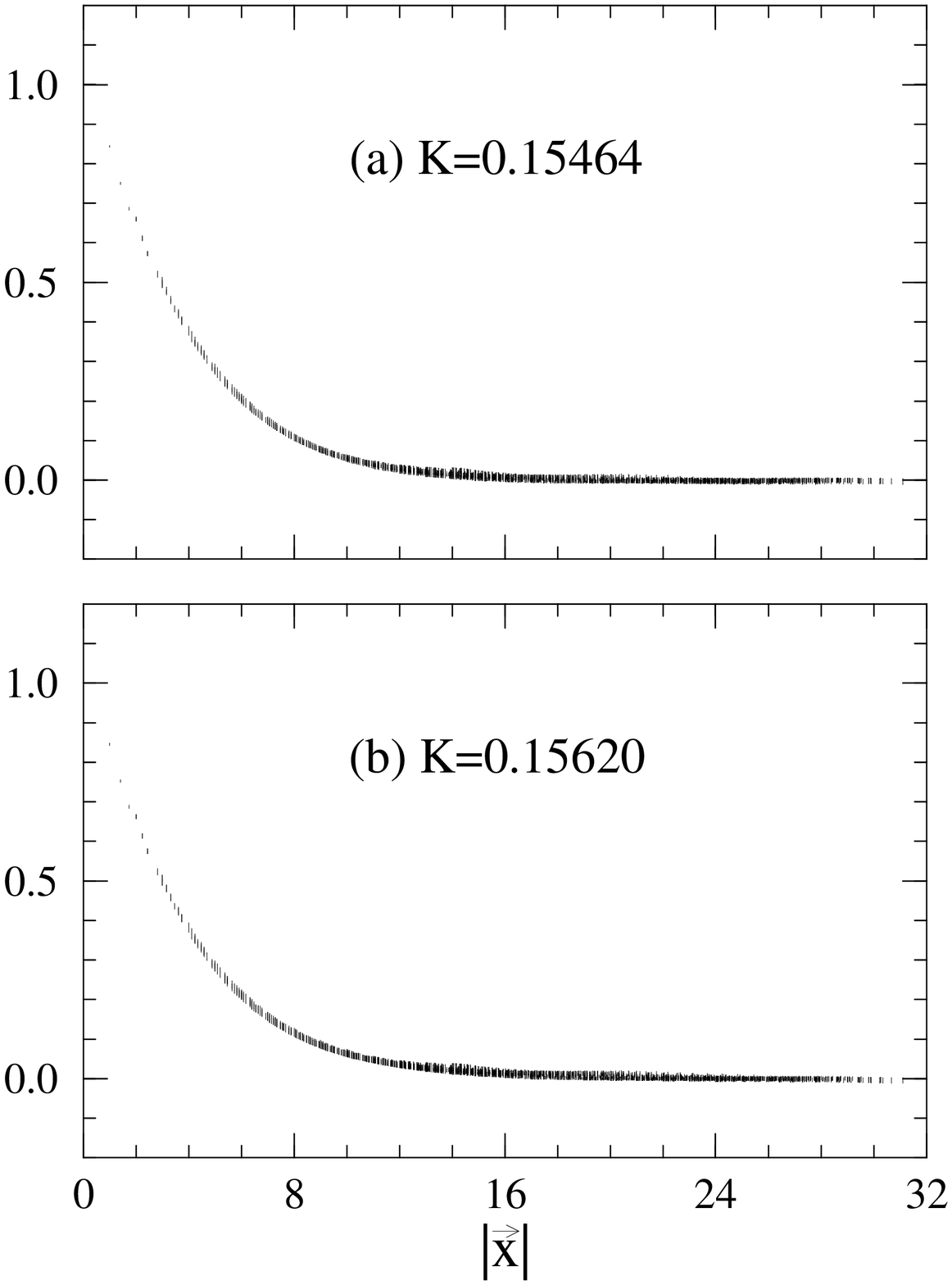,width=120mm,angle=0}
}
\caption{Quark wave function in the pion normalized
by the value at the origin for (a) $K=0.15464$ and (b) $K=0.15620$.
$|{\vec x}|$ is distance between two quarks.} 
\label{fig:wf_pi}
\end{figure}

Our calculations are carried out in three steps. 
We first measure the quark wave function in the pion for
each hopping parameter using the
ratio of eq.~(\ref{eq:wf_pi}). 
For this purpose we prepare
gauge configurations fixed to the Coulomb gauge
except the $t=0$ time slice. On these configurations
the pion correlation functions in eq.~(\ref{eq:wf_pi}) are constructed
employing the quark propagators solved with wall sources 
at the $t=0$ time slice where the Dirichlet boundary condition
is imposed in the time direction.
We note that the non-local pion sources in the $t=0$ time slice
cancel out in the average over gauge configurations.
Figure~\ref{fig:wf_pi} shows the results of $\Psi(|{\vec x}|)$
measured at $t=18$ for the heaviest ($K=0.15464$) 
and lightest ($K=0.15620$) hopping parameters.
We use the central values of  $\Psi(|{\vec x}|)$ for the smeared 
nucleon sources of eqs.~(\ref{eq:smear_p}) and (\ref{eq:smear_n}).

In the second step we calculate various two-point functions
required to determine hadron masses, $\sqrt{Z_{PS}}$,
$\sqrt{Z_{N}}$, $\alpha$ and $\beta$. 
We extract the PS meson masses and the amplitudes $\sqrt{Z_{PS}}$ 
from the correlation functions of eq.~(\ref{eq:z_ps_2pt}) where we employ
the set of quark propagators solved with the sources of 
${\rm e}^{i{\vec p}\cdot{\vec x}}$ at the $t=29$ time slice
without gauge fixing.
The nucleon masses are determined from the smeared-local correlation
function 
${\sum_{\vec x}
      \la { J}_{N,s}({\vec x},t)
          \bar{ J}^\prime_{N,s}(0)\ra}$, 
fixing gauge
configurations on the $t=0$ time slice to the Coulomb gauge.
The amplitudes $\sqrt{Z_{N}}$ are evaluated by fitting
the local-local correlation function of eq.~(\ref{eq:z_n_2pt}) 
to an exponential form 
with the nucleon mass fixed.
It is straightforward to calculate the $\alpha$ and $\beta$ parameters 
with the use of the ratio of eq.~(\ref{eq:ratio_ab}). 

\begin{figure}[h]
\centering{
\hskip -0.0cm
\psfig{file=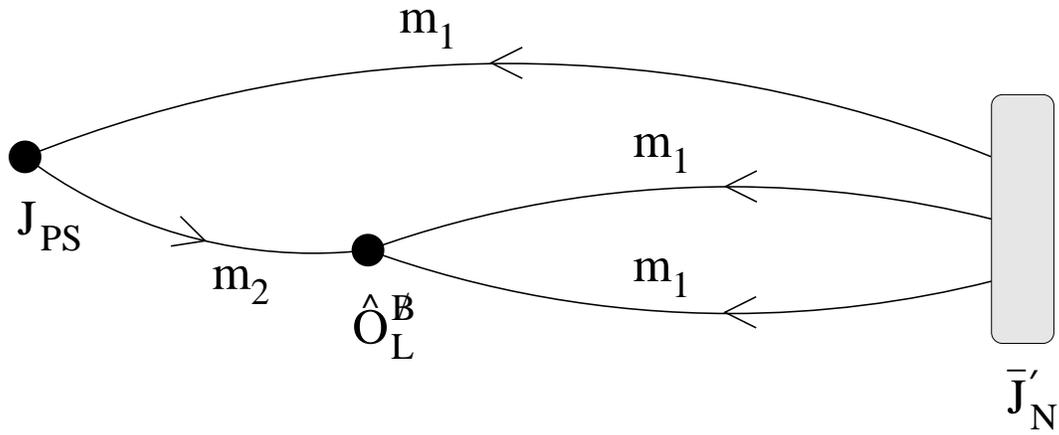,width=140mm,angle=-90}
}
\caption{Quark flow diagram for the nucleon decay three point function
with the mass assignment. Filled circles denote the local operators and
shaded rectangular is for the smeared operator.} 
\label{fig:qldgm}
\end{figure}

Finally we calculate the ratio of eq.~(\ref{eq:ratio}), where
the baryon number violating operator is moved
between the nucleon source and the PS meson sink. 
Gauge configurations on the $t=0$ time slice are fixed to the 
Coulomb gauge to employ the smeared source for the nucleon.
For the calculation of the three-point function in the 
ratio, we use the source method to insert the pion fields
at $t=29$ into the quark propagators solved with the $t=0$
smeared source\cite{source}.
We should note that calculation of the 
$p\rightarrow \eta $ matrix elements
of eq.~(\ref{eq:indme_7}) requires the disconnected diagrams
in terms of the quark lines, which cannot be calculated
by the source method. Although these diagrams could have
contributions to the matrix elements in the non-degenerate case
of the up, down and strange quark masses, 
we neglect them in this paper.
Four spatial momenta ${\vec p}a=(0,0,0), (\pi/14,0,0), (0,\pi/14,0)$
and $(0,0,\pi/24)$ are imposed on the PS meson in the final
state. For the ${\vec p}\ne {\vec 0}$ cases
we distinguish the strange quark mass from the
up and down quark mass by providing different hopping parameters
for $m_1$ and $m_2$ in Fig.~\ref{fig:qldgm}. 
As explained in Sec.~\ref{subsec:ff}, we cannot disentangle
the relevant form factor from the irrelevant one
in the case of the PS meson at rest, where we take
only the degenerate quark mass $m_1=m_2$.

From the tree-level expressions of chiral lagrangian 
for the nucleon decay matrix elements
in eqs.~(\ref{eq:chpt_1_rl})$-$(\ref{eq:chpt_7_ll}), 
we can assume that
the form factors obtained from the ratio  of eq.~(\ref{eq:ratio})
are functions of $q^2$, $m_1$ and $m_2$, where the $m_1$ and $m_2$
dependences could appear through the baryon masses, the pion decay constant
and the $\alpha$, $\beta$, $F$, $D$, $b_{1,2}$ parameters,
To interpolate the form factors to the $q^2=0$ point,  
where the charged lepton masses are negligible 
(see Sec.~\ref{sec:ndecay}),
we employ the following fitting function
\be
c_0+c_1\cdot(-q^2) +c_2\cdot (-q^2)^2 +c_3\cdot m_1 +c_4\cdot m_2.
\label{eq:qfit}
\ee
We extrapolate  $m_1$ and $m_2$ to the chiral limit
for the matrix elements of 
eqs.~(\ref{eq:indme_1}), (\ref{eq:indme_2}) and (\ref{eq:indme_7}), while
$m_2$ is interpolated to the physical strange quark mass
with $m_1$ taken to the chiral limit  
for the matrix elements of eqs.~(\ref{eq:indme_3})$-$(\ref{eq:indme_6}).

To calculate the perturbative renormalization factors, we
determine the strong coupling constant at the scale $1/a$ and $\pi/a$ 
in the ${\overline {\rm MS}}$ scheme.
We first define the coupling constant $\alpha_P$\cite{alpha_p} 
from the expectation value of the
plaquette $P=\la {\rm Tr}U_P\ra/3$:
\be 
-{\rm ln}P=\frac{4\pi}{3}\alpha_P(3.40/a)
\left[1-1.19\alpha_P\right].
\ee
The conversion from $\alpha_P$ to the ${\overline {\rm MS}}$ coupling
constant is made by
\be
\alpha_{\overline {\rm MS}}(3.40/a)=\alpha_P({\rm e}^{\frac{5}{6}}3.40/a)
\left[1+\frac{2}{\pi}\alpha_P+0.95\alpha_P^2\right]. 
\ee
The values of $\alpha_{\overline {\rm MS}}(1/a)$ and 
 $\alpha_{\overline {\rm MS}}(\pi/a)$ are obtained by two-loop
renormalization group running starting from 
$\alpha_{\overline {\rm MS}}(3.40/a)$.

We estimate errors by the single elimination jackknife procedure for all
measured quantities.

\section{Results for $\alpha$ and $\beta$ parameters}
\label{sec:ab}

\begin{figure}[h]
\centering{
\hskip -0.0cm
\psfig{file=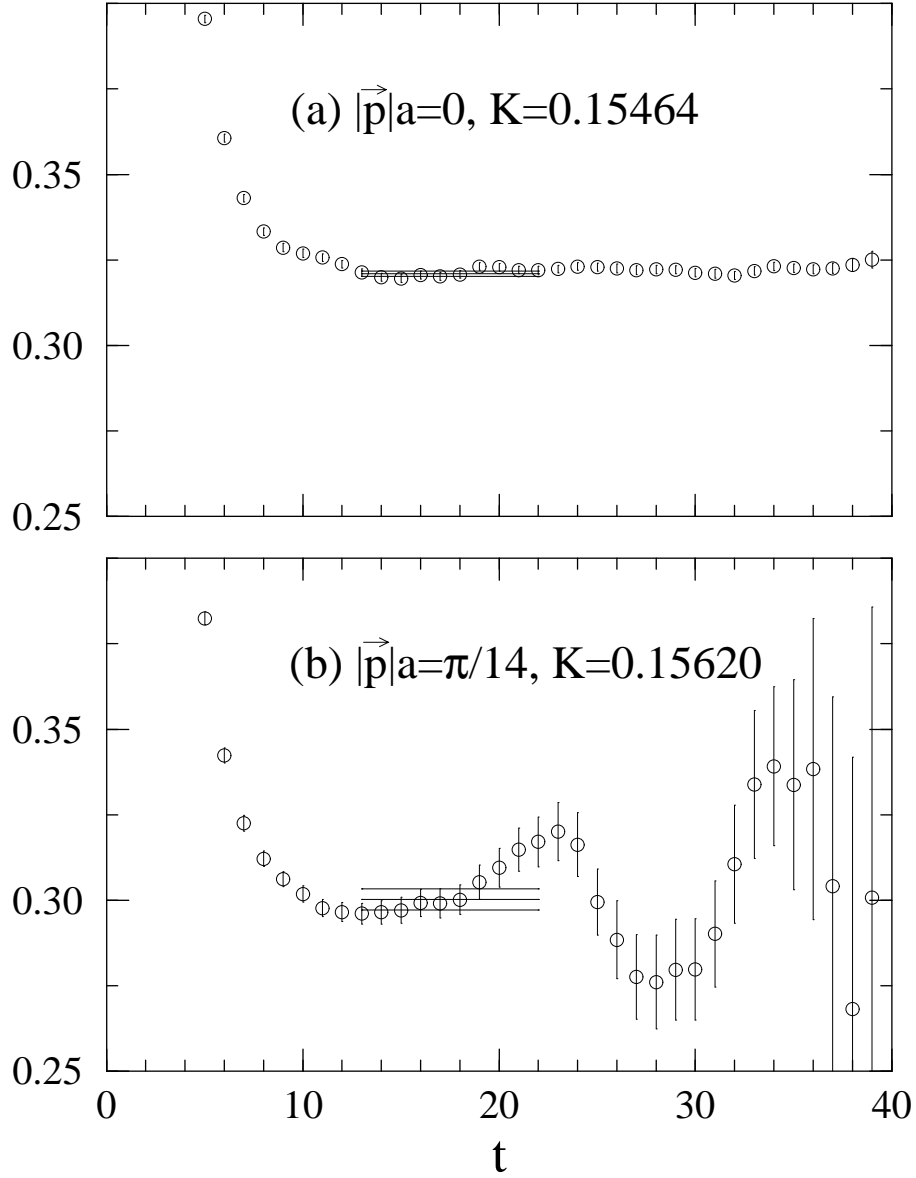,width=120mm,angle=0}
}
\caption{(a) Effective mass for the pion with $|{\vec p}|a=0$ 
at $K=0.15464$ 
and (b) effective energy for the pion with  $|{\vec p}|a=\pi/14$ 
at $K=0.15620$.
The pion correlation functions consist of the local sink and the wall source
without gauge fixing.
Solid lines denote the fitting results with an error band  of one standard 
deviation obtained by global fits of the pion propagators.} 
\label{fig:eff_ps_w}
\end{figure}

In this section we present the results for hadron masses, 
$\sqrt{Z_{PS}}$, $\sqrt{Z_{N}}$, $\alpha$ and $\beta$ which are
obtained from the two-point functions.
In Fig.~\ref{fig:eff_ps_w} we plot effective masses
of the PS meson for the case of $|{\vec p}|a=0$ at $K=0.15464$ and 
$|{\vec p}|a=\pi/14$ at $K=0.15620$; the statistical errors are
best controlled in the former and worst in the latter.
We observe plateaus beyond $t\approx 13$ for both cases.
The horizontal lines denote the fitted values
of the PS meson masses with an error of  one standard deviation
obtained by a global fit of
the two-point function of eq.~(\ref{eq:z_ps_2pt})
with the function
\be
\frac{Z_{PS}}{2m_{PS}}\left( {\rm e}^{-m_{PS}(t-29)}
+ {\rm e}^{-m_{PS}(T-t+29)}\right),
\ee 
where the fitting range is chosen to be  $13\le t\le 22$ after
taking account of the time reversal symmetry 
$t-29 \leftrightarrow T-t+29$.    
This fitting procedure also gives the amplitude $Z_{PS}$.
We tabulate the numerical values of $m_{PS}$ 
in Tables~\ref{tab:hmass} and those of the PS meson energy for the
case of $|{\vec p}|\ne 0$ in Table~\ref{tab:q2a2}.

\begin{figure}[h]
\centering{
\hskip -0.0cm
\psfig{file=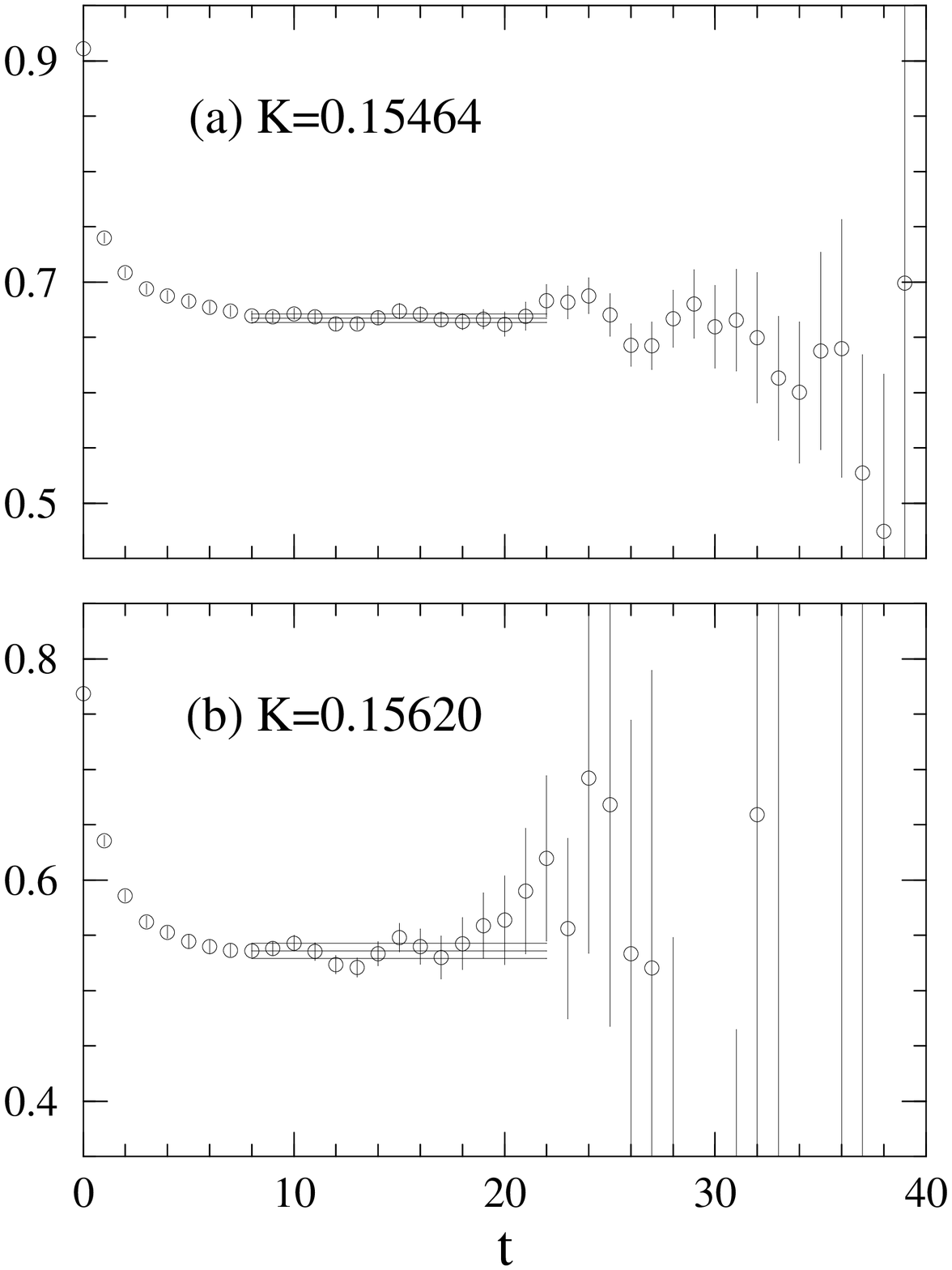,width=120mm,angle=0}
}
\caption{Effective mass for the nucleon with the smeared source
at (a) $K=0.15464$ and (b) $K=0.15620$.
Solid lines denote the fitting results with an error band  of one standard 
deviation obtained by global fits of the nucleon smeared-local propagators.} 
\label{fig:eff_n_sp0}
\end{figure}

\begin{figure}[h]
\centering{
\hskip -0.0cm
\psfig{file=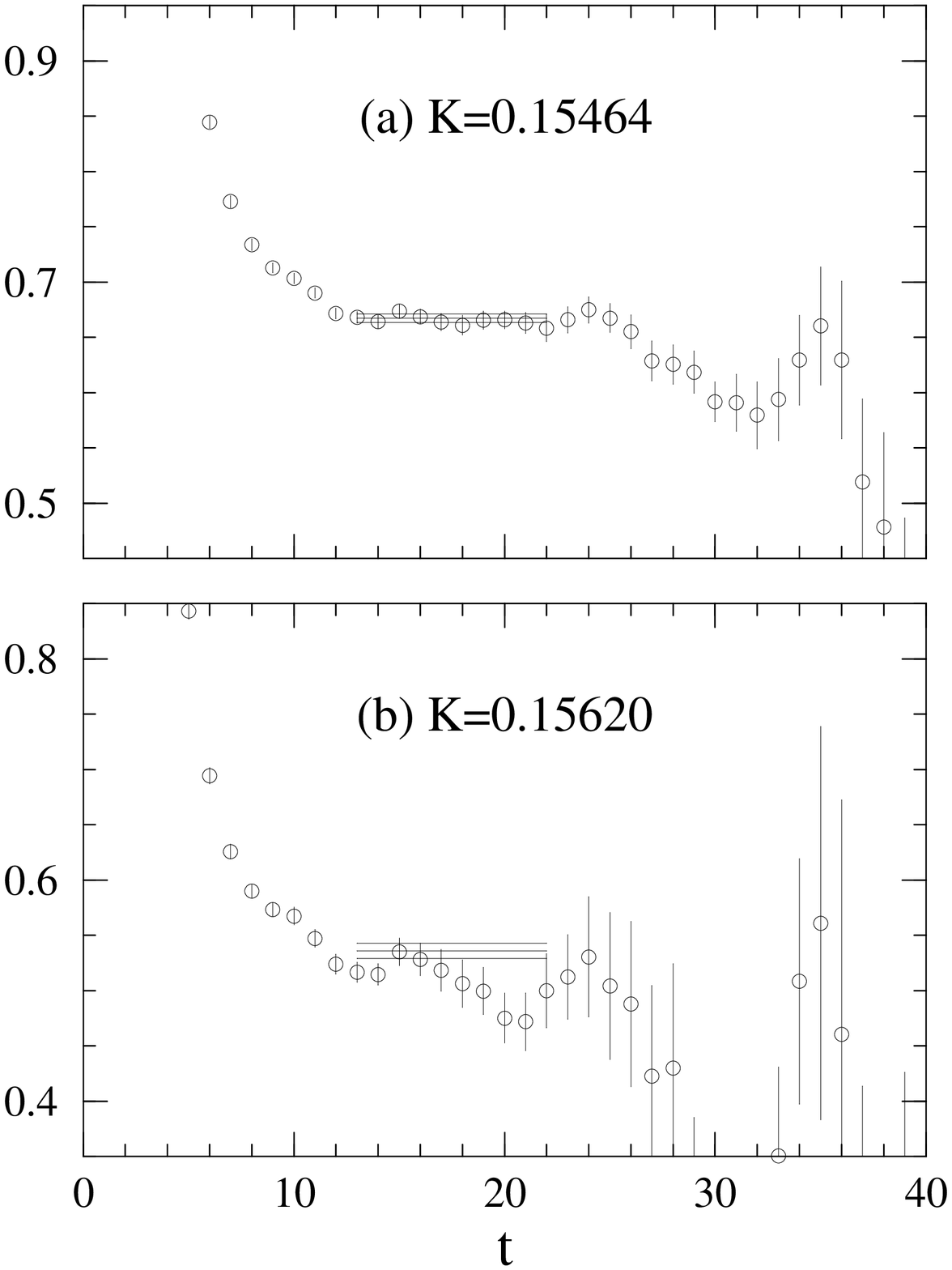,width=120mm,angle=0}
}
\caption{Effective mass for the nucleon with the local source
at (a) $K=0.15464$ and (b) $K=0.15620$.
Solid lines denote the fitting results with an error band  of one standard 
deviation obtained by global fits of the nucleon smeared-local propagators.} 
\label{fig:eff_n_lp0}
\end{figure}

Figure~\ref{fig:eff_n_sp0} shows the nucleon effective masses
obtained from the smeared-local correlation functions for
the heaviest($K=0.15464$) and lightest($K=0.15620$) quark masses, which
should be compared with Fig.~\ref{fig:eff_n_lp0} for the local-local
correlation functions.
We observe that the smeared source works effectively,
dominantly overlapping with the lowest energy state.
We extract the nucleon masses by fitting the smeared-local
correlation functions to a exponential form 
with the
fitting range $8 \le t \le 22$. The fitted values are
shown by the horizontal lines in Figs.~\ref{fig:eff_n_sp0} and
\ref{fig:eff_n_lp0} together with one standard deviation errors. 
The amplitudes $Z_N$ defined in eq.~(\ref{eq:z_n}) 
are obtained by a fit of the local-local
correlation functions with the function
\be
Z_N {\rm e}^{-m_N t}
\ee
over the range $13\le t \le 22$,
where $m_N$ is fixed to be the value 
determined from the smeared-local correlation
functions.
We present the numerical values of $m_N$ 
for the four hopping parameters in Table~\ref{tab:hmass}.

\begin{figure}[h]
\centering{
\hskip -0.0cm
\psfig{file=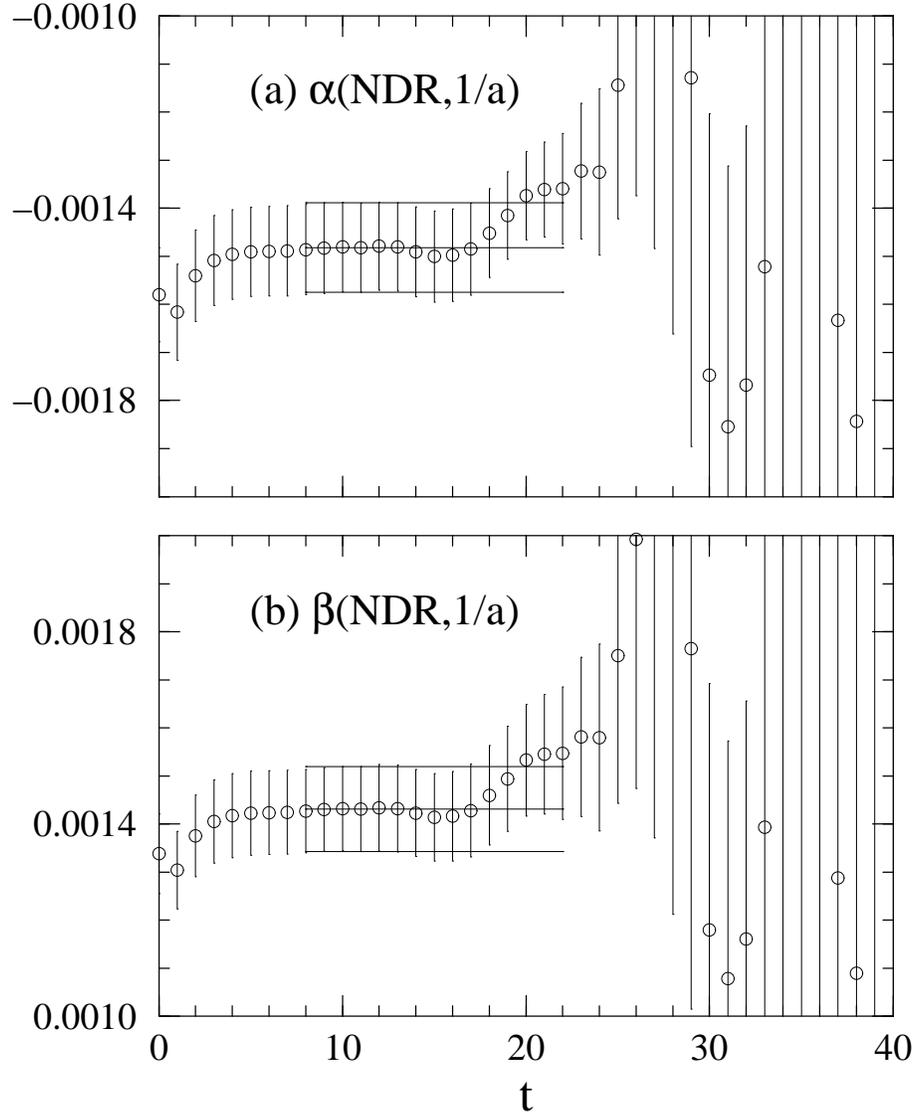,width=120mm,angle=0}
}
\caption{Ratio $R^{\alpha\beta}(t)$ for
(a) $\alpha$ and (b) $\beta$ parameters at $K=0.15620$.
Solid lines denote the fitting results with an error band  of one standard 
deviation.} 
\label{fig:ratio_ab}
\end{figure}

\begin{figure}[h]
\centering{
\hskip -0.0cm
\psfig{file=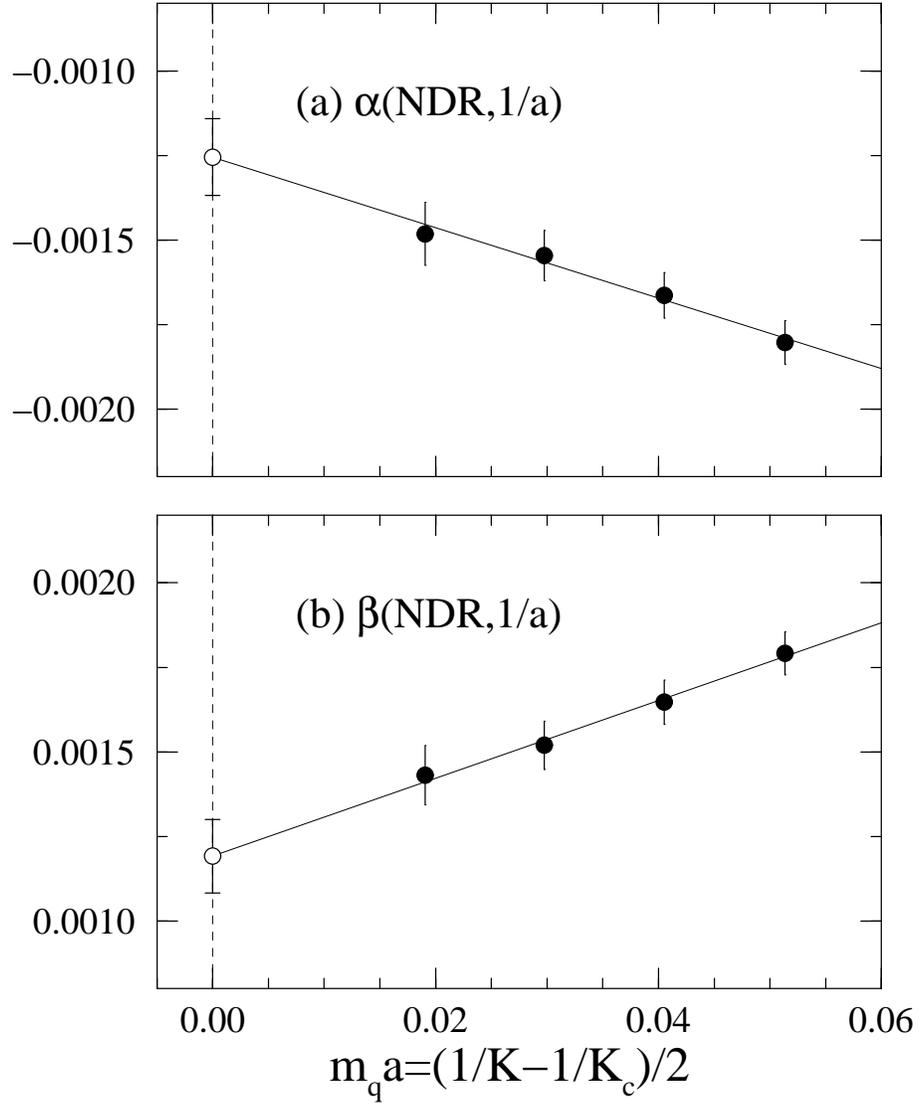,width=120mm,angle=0}
}
\caption{Chiral extrapolations of
(a) $\alpha$ and (b) $\beta$ parameters.
Solid lines denote linear fits.} 
\label{fig:ab_chl}
\end{figure}

The $\alpha$ and $\beta$ parameters are extracted from a
constant fit of the ratio of eq.~(\ref{eq:ratio_ab}), which 
is shown in Fig.~\ref{fig:ratio_ab} 
for the case of the lightest quark mass($K=0.15620$).
The horizontal lines represent the fit 
with the fitting range chosen to be $8 \le t \le 22$.
The numerical values are given in Table~\ref{tab:ab}.
Figure~\ref{fig:ab_chl} illustrates quark mass dependences
of the $\alpha$ and $\beta$ parameters.
Applying linear fits to the data, we obtain
$\alpha({\rm NDR},1/a)=-0.015(1)$GeV$^3$ and 
$\beta({\rm NDR},1/a)=0.014(1)$GeV$^3$ 
in the chiral limit with the use of $a^{-1}=2.30(4)$GeV. 

Let us compare our results for the  $\alpha$ and $\beta$ parameters
with the previous estimates.
We summarize the previous lattice results 
in Table~\ref{tab:ab_latt} together with
the simulation parameters. 
In Refs.~\cite{bowler,gavela} the lattice cut-off scale $a^{-1}$ was 
determined by the nucleon mass.  
The nucleon mass results employed \cite{bowler_spc,ms_spc} 
are, however, quite heavy compared to 
those of more recent high statistical calculations\cite{gf11,gupta}:
$m_Na=1.11(10)$\cite{bowler_spc} compared to $m_Na=0.756(19)$\cite{gf11} 
in the chiral limit at $\beta=5.7$, and  
$m_Na=0.64(11)$\cite{ms_spc} compared to $m_Na=0.461(9)$\cite{gupta} 
in the chiral limit at $\beta=6.0$. 
To avoid this large uncertainty,
we employ $a^{-1}$ determined by the $\rho$ meson mass 
to obtain the $\alpha$ and $\beta$ parameters in physical units
in Table~\ref{tab:ab_latt}.

In phenomenological GUT model analyses of the nucleon decays, 
the values $|\alpha|=|\beta|=0.003$GeV$^3$\cite{ab_min} are conservatively 
taken as these are the smallest estimate among various
QCD model calculations\cite{model_wf}.  
A trend one observes in Table~\ref{tab:ab_latt} is that the previous 
lattice calculations indicated values of these parameters 
considerably larger than the minimum model estimate above. 
Our results, significantly improved over the previous ones 
due to the use of higher statistics, larger spatial size, 
lighter quark masses and smaller lattice spacing, 
have confirmed this trend:
the values we obtained are about five times larger than 
$|\alpha|=|\beta|=0.003$GeV$^3$.

\section{Results for nucleon decay matrix elements}
\label{sec:ndecay}

\begin{figure}[h]
\centering{
\hskip -0.0cm
\psfig{file=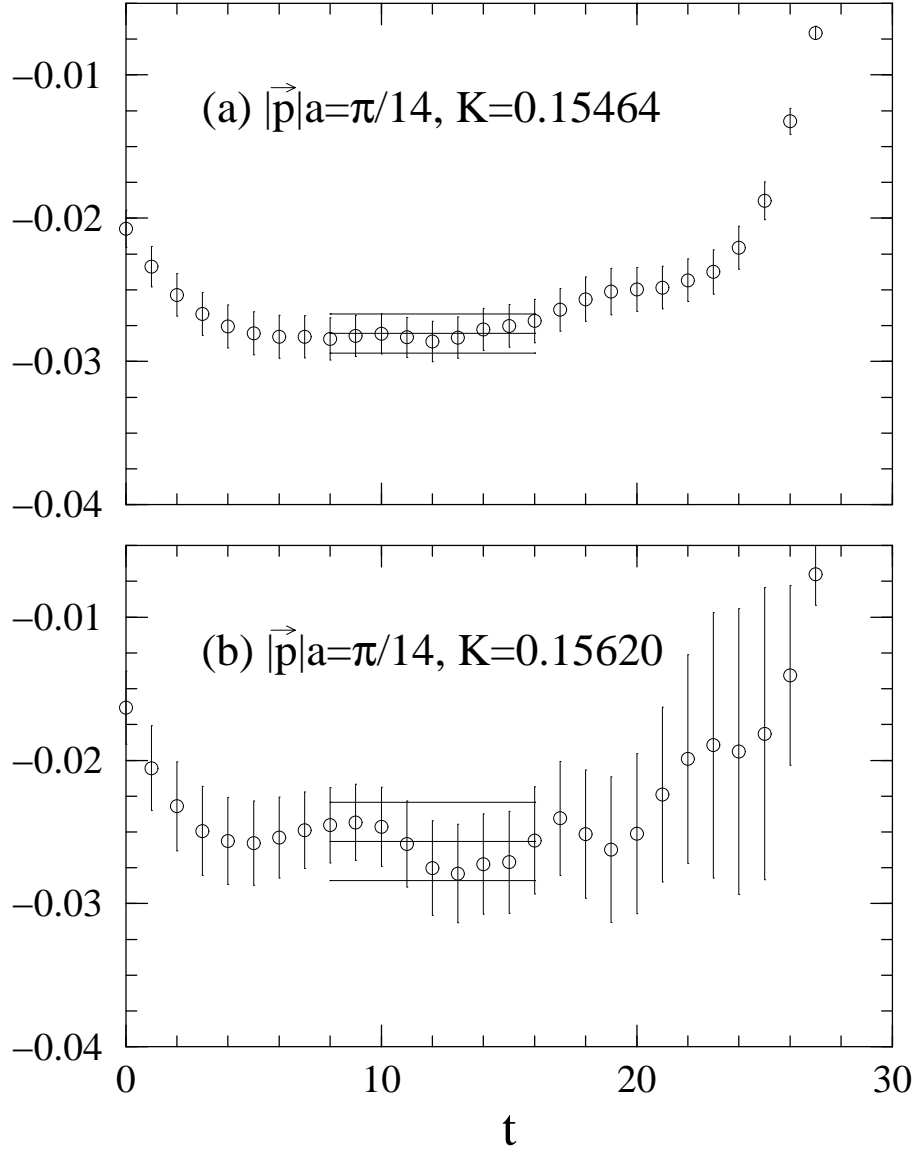,width=120mm,angle=0}
}
\caption{Ratio $R(t,t^\prime=29)$ for the relevant form factor
in $\langle \pi^0|(ud_R) u_L|p\rangle$ 
at (a) $K=0.15464$ and (b) $K=0.15620$.
Solid lines denote the fitting results with an error band of  one standard
deviation.} 
\label{fig:ratio_3pt_xy}
\end{figure}

We now turn to the calculation of the nucleon matrix elements
with the direct method. In Fig.~\ref{fig:ratio_3pt_xy} we show time
dependences of $R(t,t^\prime=29)$ with $|{\vec p}|a=\pi/14$ for
the matrix element $\la \pi^0|\epsilon_{ijk}
({u^i}^T CP_{R}d^j) P_L u^k|p\ra$ 
in the case of the heaviest quark mass($K=0.15464$) 
and the lightest one($K=0.15620$). The results of constant fits
are represented by the sets of three  horizontal lines.
We choose the fitting range to be $8 \le t \le 16$
for all the matrix elements of 
eqs.~(\ref{eq:indme_1})$-$(\ref{eq:indme_7}) such that
the excited state contaminations in the nucleon
and PS meson states observed in Figs.\ref{fig:eff_ps_w} and 
\ref{fig:eff_n_sp0} can be avoided simultaneously.

\begin{figure}[h]
\centering{
\hskip -0.0cm
\psfig{file=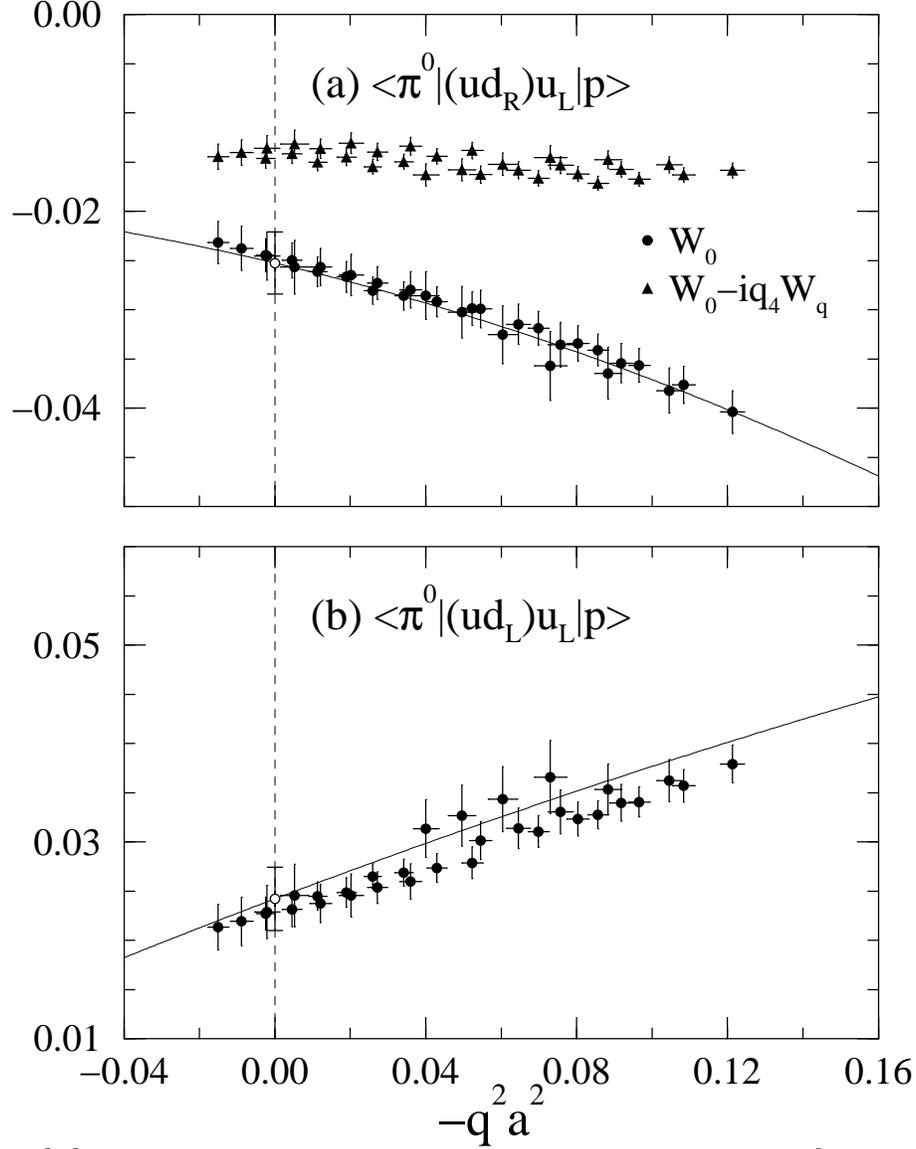,width=120mm,angle=0}
}
\caption{$-q^2 a^2$ dependences for the relevant form factor $W_0$
in (a) $\langle \pi^0|(ud_R) u_L|p\rangle$ 
and (b) $\langle \pi^0|(ud_L) u_L|p\rangle$.
Combination of form factors $W_0-iq_4 W_q$ 
is also plotted in (a) for comparison. 
Solid lines denote the function 
$c_0+c_1\cdot (-q^2 a^2)+c_2\cdot (-q^2a^2)^2$.} 
\label{fig:qfit1}
\end{figure}

\begin{figure}[h]
\centering{
\hskip -0.0cm
\psfig{file=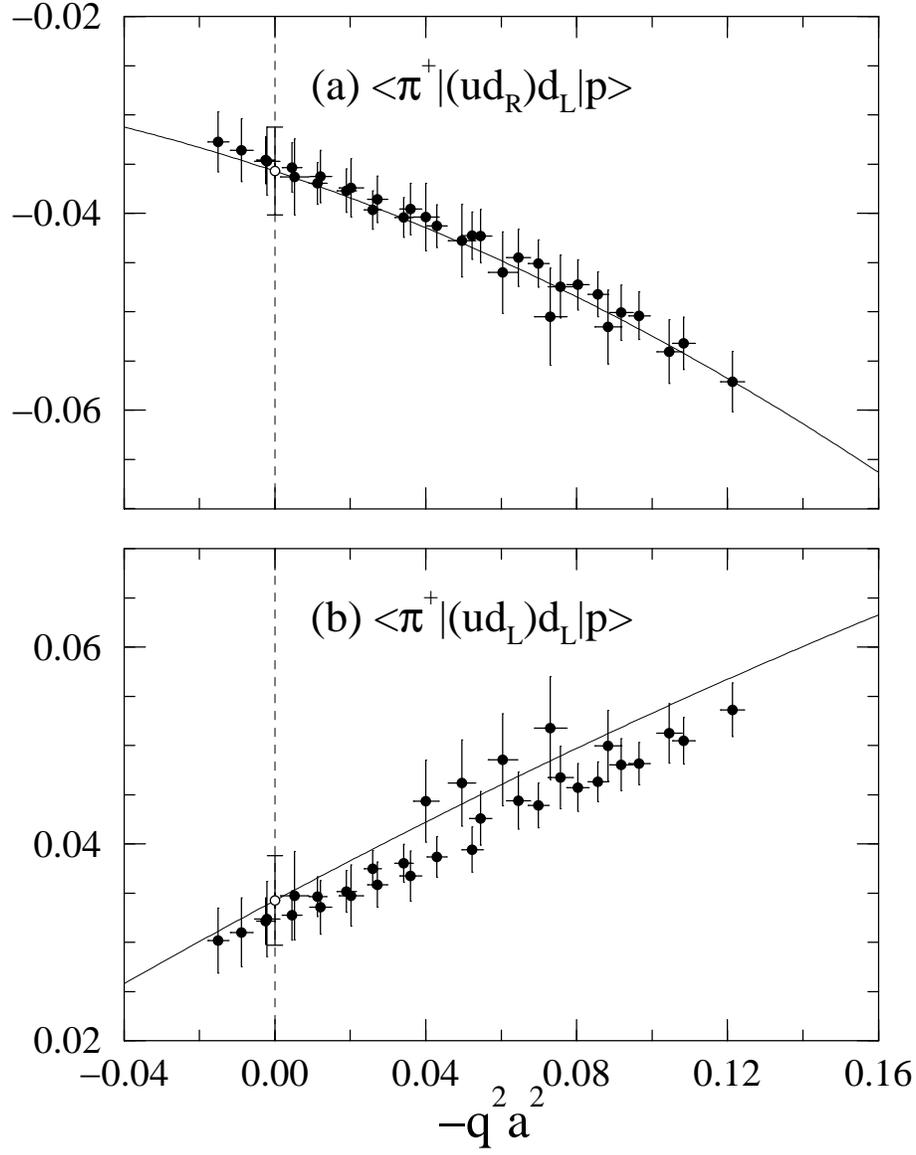,width=120mm,angle=0}
}
\caption{Same as Fig.~\protect{\ref{fig:qfit1}}
for (a) $\langle \pi^+|(ud_R) d_L|p\rangle$ 
and (b) $\langle \pi^+|(ud_L) d_L|p\rangle$.} 
\label{fig:qfit2}
\end{figure}

\begin{figure}[h]
\centering{
\hskip -0.0cm
\psfig{file=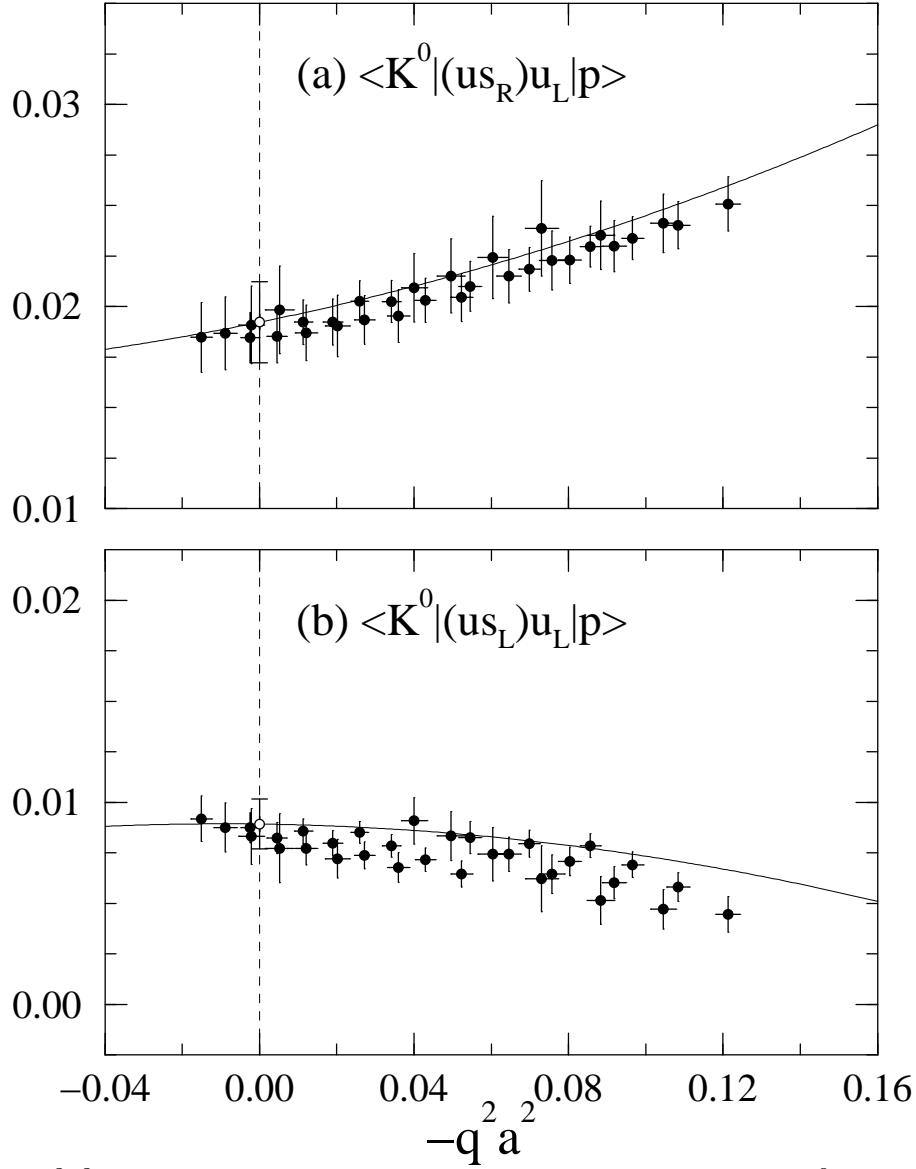,width=120mm,angle=0}
}
\caption{$-q^2 a^2$ dependences for the relevant form factor $W_0$
in (a) $\langle K^0|(us_R) u_L|p\rangle$ 
and (b) $\langle K^0|(us_L) u_L|p\rangle$.
Solid lines denote the function 
$c_0+c_1\cdot (-q^2 a^2)+c_2\cdot (-q^2a^2)^2+c_4\cdot m_s a$.} 
\label{fig:qfit3}
\end{figure}

\begin{figure}[h]
\centering{
\hskip -0.0cm
\psfig{file=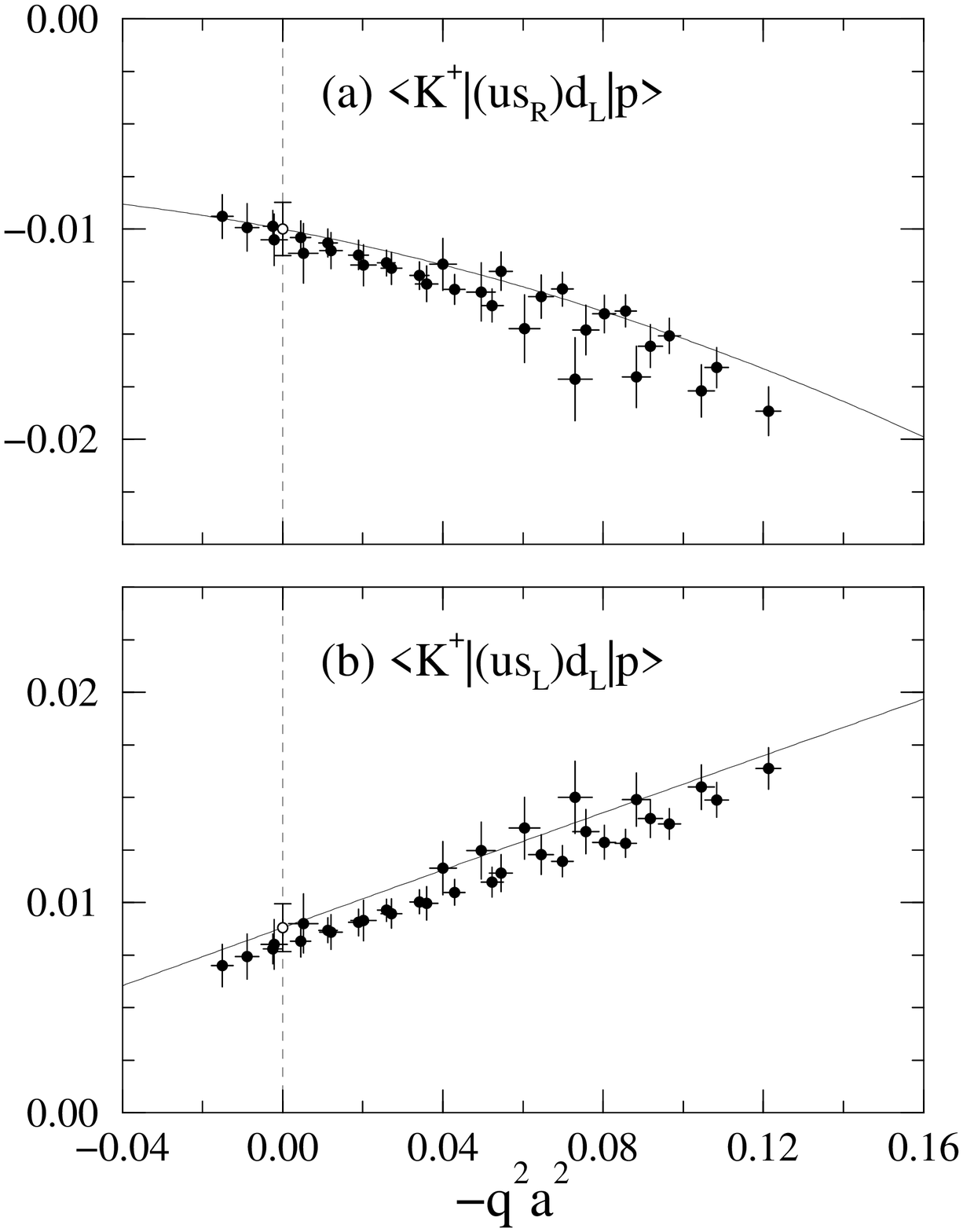,width=120mm,angle=0}
}
\caption{Same as Fig.~\protect{\ref{fig:qfit3}}
for (a) $\langle K^+|(us_R) d_L|p\rangle$ 
and (b) $\langle K^+|(us_L) d_L|p\rangle$.}
\label{fig:qfit4}
\end{figure}

\begin{figure}[h]
\centering{
\hskip -0.0cm
\psfig{file=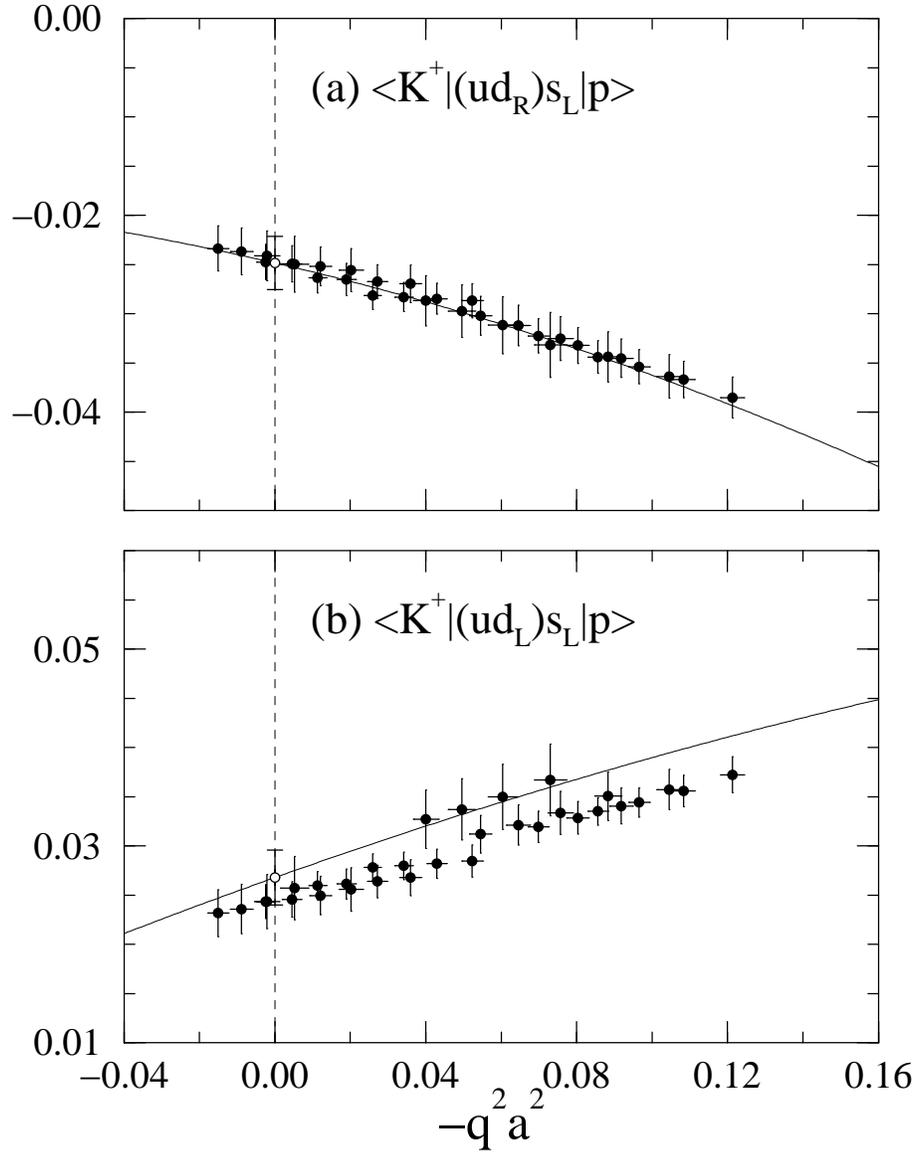,width=120mm,angle=0}
}
\caption{Same as Fig.~\protect{\ref{fig:qfit3}}
for (a) $\langle K^+|(ud_R) s_L|p\rangle$ 
and (b) $\langle K^+|(ud_L) s_L|p\rangle$.}
\label{fig:qfit5}
\end{figure}

\begin{figure}[h]
\centering{
\hskip -0.0cm
\psfig{file=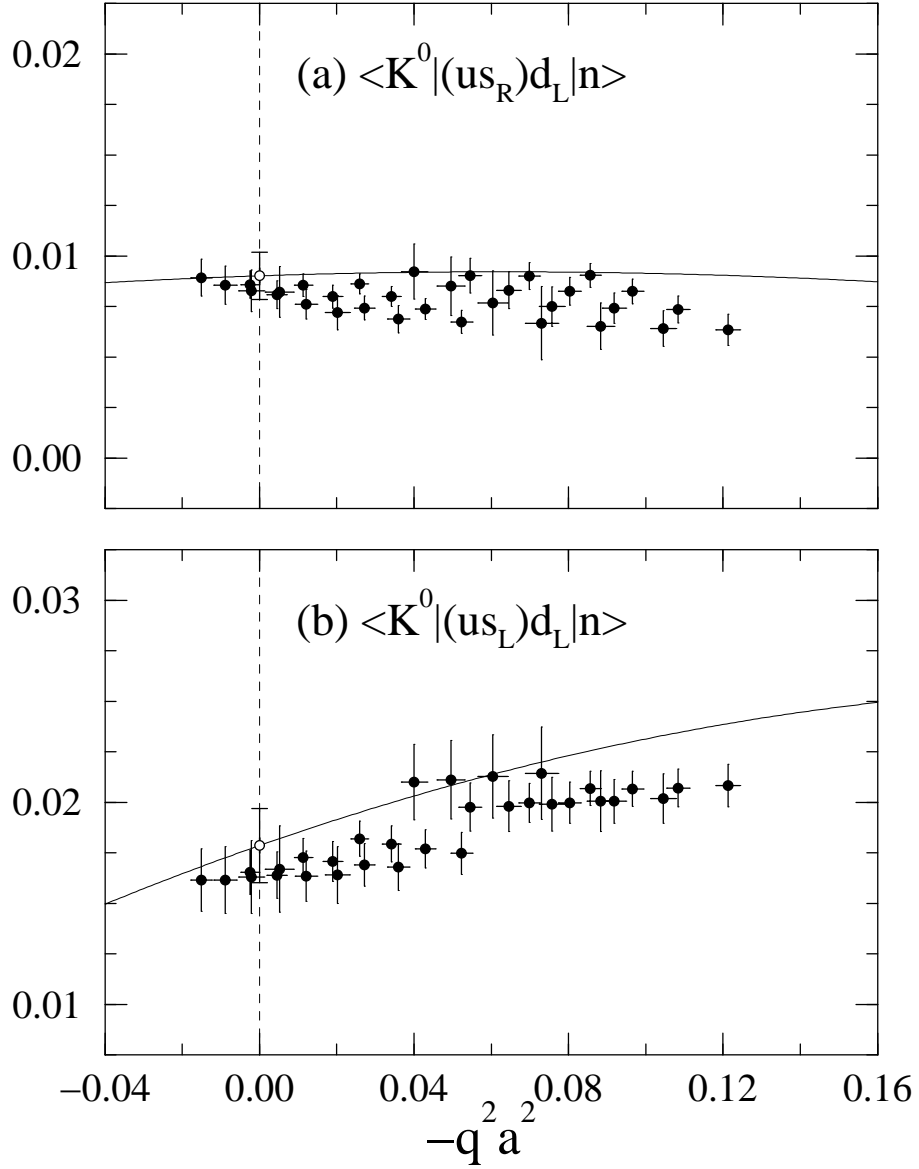,width=120mm,angle=0}
}
\caption{Same as Fig.~\protect{\ref{fig:qfit3}}
for (a) $\langle K^0|(us_R) d_L|n\rangle$ 
and (b) $\langle K^0|(us_L) d_L|n\rangle$.}
\label{fig:qfit6}
\end{figure}

\begin{figure}[h]
\centering{
\hskip -0.0cm
\psfig{file=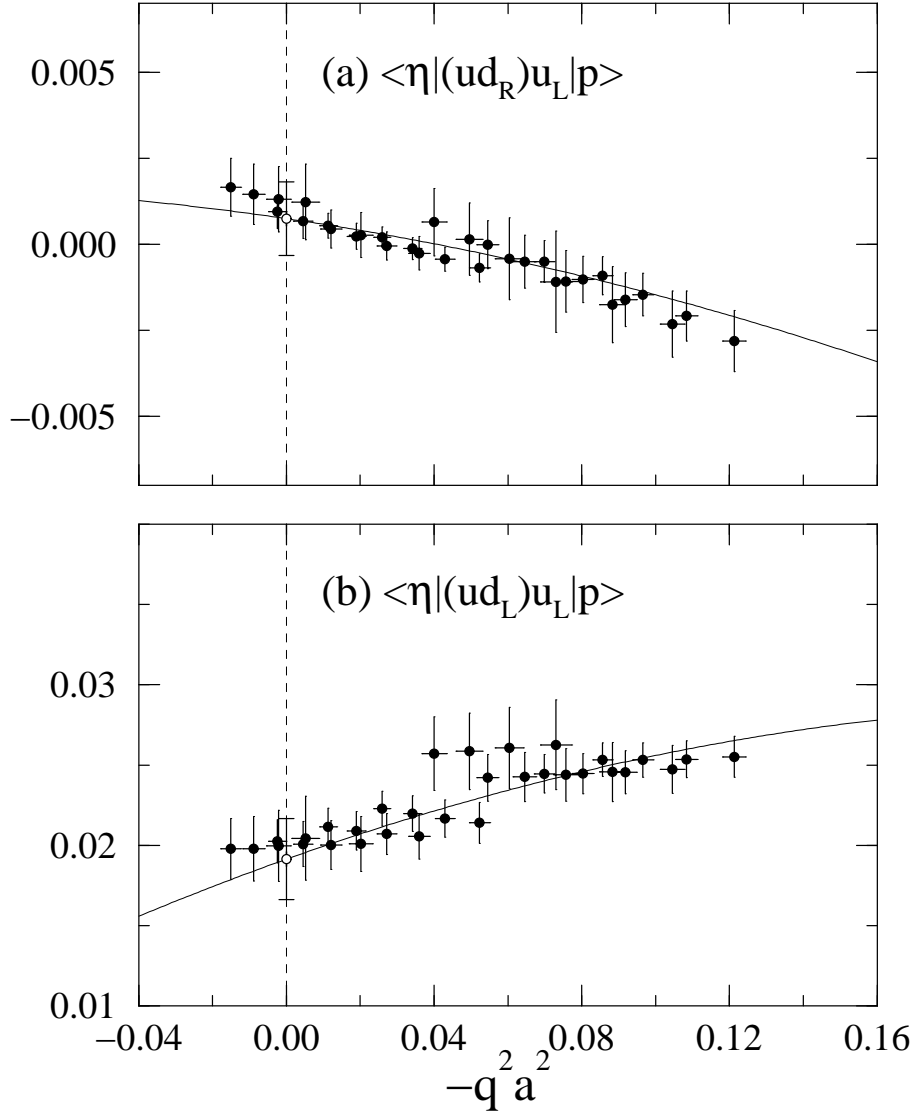,width=120mm,angle=0}
}
\caption{Same as Fig.~\protect{\ref{fig:qfit1}}
for (a) $\langle \eta |(ud_R) u_L|p\rangle$ 
and (b) $\langle \eta |(ud_L) u_L|p\rangle$.}
\label{fig:qfit7}
\end{figure}

Figures~\ref{fig:qfit1}$-$\ref{fig:qfit7} show $-q^2a^2$ dependences
of the relevant form factors $W_0(q^2)$ in the independent 
nucleon decay matrix elements
in eqs.~(\ref{eq:indme_1})$-$(\ref{eq:indme_7}), where 
the operators are renormalized with the NDR scheme at $\mu=1/a$.
The values of $-q^2 a^2$ 
are enumerated in Table~\ref{tab:q2a2}
as a function of the quark masses $m_{1,2}$ and the 
spatial momentum ${\vec p}$.
In Fig.~\ref{fig:qfit1} (a)
we also plotted the combination $W_0-iq_4 W_q$ for comparison, which
is obtained by following the method in Ref.~\cite{gavela}.
The magnitude of  $W_0(q^2)$ is more than two times larger than that
of $W_0(q^2)-iq_4 W_q(q^2)$.
The relevant form factors at $-q^2a^2=0$(open circles) 
in Figs.~\ref{fig:qfit1}$-$\ref{fig:qfit7}
are obtained by fitting
the data employing 
the function of eq.~(\ref{eq:qfit}), where we find that
the charged lepton masses $m_e^2 a^2=4.9\times 10^{-8}$ 
and $m_\mu^2 a^2=2.1\times 10^{-3}$ 
are negligible in the current numerical statistics.
We plot the function 
$c_0+c_1\cdot(-q^2)+c_2\cdot(-q^2)^2$ 
employing the fitting results of $c_0$, $c_1$ and $c_2$
in Figs.~\ref{fig:qfit1}, \ref{fig:qfit2} and \ref{fig:qfit7},
and $c_0+c_1\cdot(-q^2)+c_2\cdot(-q^2)^2+c_4\cdot m_s$ 
with the fitting results of  $c_0$, $c_1$, $c_2$ and $c_4$
in Figs.\ref{fig:qfit3}$-$\ref{fig:qfit6}.
We observe that the signs of $c_0$ and $c_1$ are consistent with
the predictions of chiral lagrangian
in eqs.~(\ref{eq:chpt_1_rl_q})$-$(\ref{eq:chpt_7_ll_q}) 
for all the matrix elements, while the signs of $c_2$ show
disagreement in some matrix elements. 
The coefficients $c_2$, however, are poorly
determined compared to $c_0$ and $c_1$.
The fitting results for $W_0(q^2=0)$ are presented
in Table~\ref{tab:summary}.

\begin{figure}[h]
\centering{
\hskip -0.0cm
\psfig{file=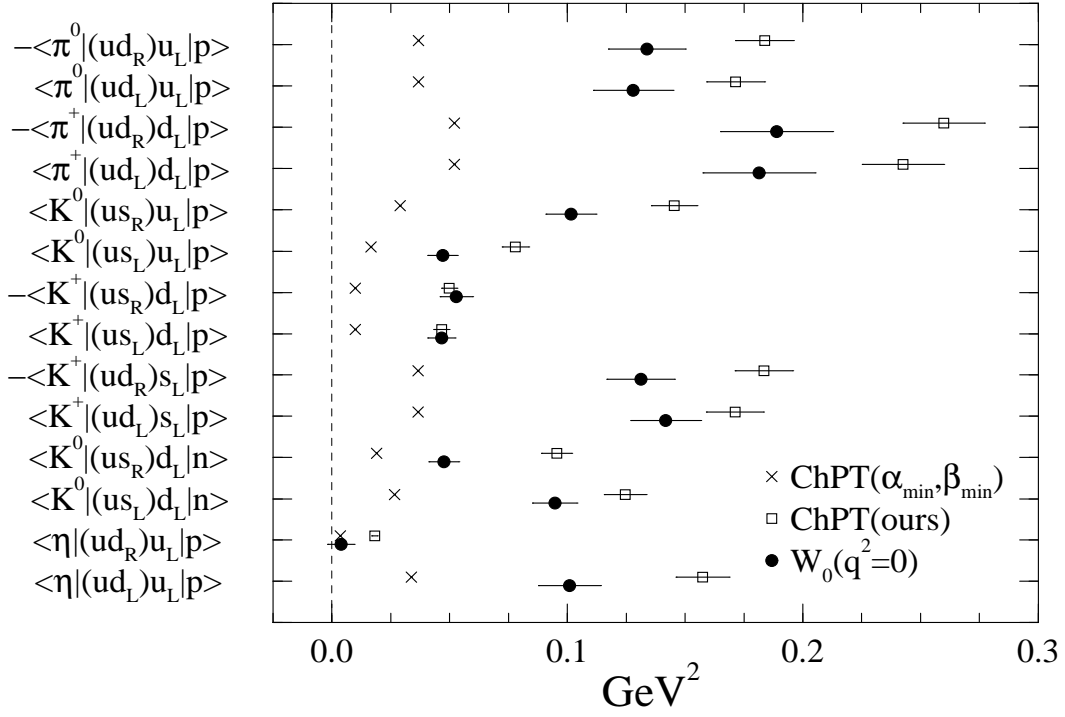,width=140mm,angle=-90}
}
\caption{Comparison of relevant form factors with tree-level predictions
of ChPT. Crosses denote the ChPT results with
$|\alpha|=|\beta|=0.003$GeV$^3$.
} 
\label{fig:summary}
\end{figure}

In Fig.~\ref{fig:summary} we compare the nucleon decay matrix
elements obtained by the direct
method with those by the indirect one using the tree-level results
of chiral lagrangian (squares), where we employ the expressions of
eqs.(\ref{eq:chpt_1_rl_q})$-$(\ref{eq:chpt_7_ll_q}) 
with $\alpha({\rm NDR},1/a)=-0.015(1)$GeV$^3$,
$\beta({\rm NDR},1/a)=0.014(1)$GeV$^3$, $f_\pi=0.131$GeV, $m_N=0.94$GeV,
$m_B=1.15$GeV, $D=0.80$ and $F=0.47$.
We observe that the two set of results are roughly comparable.  
This leads us to consider that the large discrepancy between the results 
of the two methods found in Refs.~\cite{gavela,jlqcd_98} 
is mainly due to the neglect of the
$W_q(q^2)$ term in eq.~(\ref{eq:ff}).
  
It is also intriguing to compare our results with 
the tree-level predictions of chiral lagrangian 
with $|\alpha|=|\beta|=0.003$GeV$^3$ (crosses)
that is the smallest estimate among various
QCD model calculations.
Our results with the direct method are $3-5$ times larger than the
smallest estimates except $\langle \eta |(ud_R) u_L|p\rangle$.  
Hence they are expected to give stronger 
constraints on the parameters of GUT models. 

\begin{figure}[h]
\centering{
\hskip -0.0cm
\psfig{file=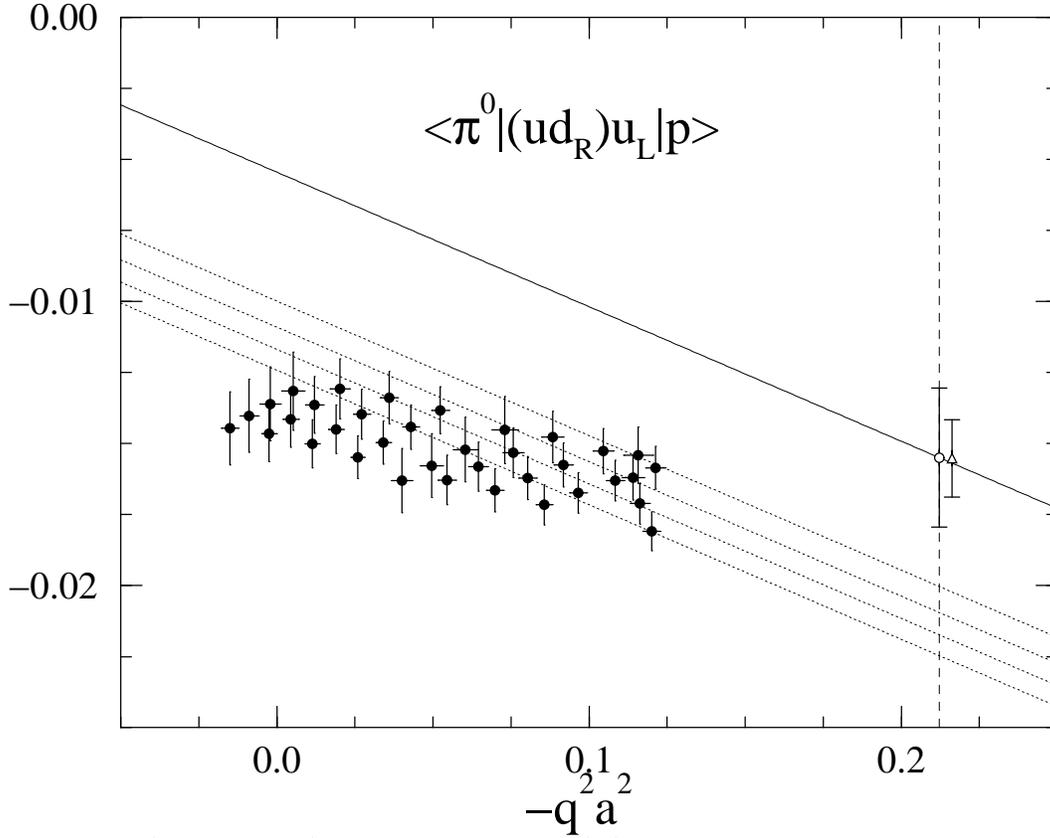,width=140mm,angle=-90}
}
\caption{$W_0(q^2)-iq_4 W_q(q^2)$ as a function of $-q^2a^2$.
Dashed vertical line denote the soft pion limit 
$-q^2 a^2=m_N^2 a^2=(0.4607)^2$.
See text for solid and dotted lines.
Triangle denotes the results for $\alpha /(\surd 2f)$.} 
\label{fig:softp}
\end{figure}

Finally, let us discuss the soft pion limit of the nucleon decay
matrix elements. The tree-level result of the chiral lagrangian
for $\langle \pi^0|(ud_R) u_L|p\rangle$ 
in eq.~(\ref{eq:chpt_1_rl}) shows that the combination
of form factors $W_0(q^2)-iq_4 W_q(q^2)$ converges to a finite value of
$\alpha/(\sqrt{2}f)$
in the soft pion limit $p_\mu \rightarrow 0$ $(-q^2=m_N^2)$,
whereas each of $W_0$ and $W_q$ diverges.
In Fig.~\ref{fig:softp} we plot $W_0-iq_4 W_q$ for 
the matrix element $\langle \pi^0|(ud_R) u_L|p\rangle$
as a function of $-q^2 a^2$. In this case the results for the pion 
at rest are also included. To extrapolate the data
to the point $-q^2 a^2=m_N^2 a^2$ (dashed vertical line),
we employ the fitting function
\be
c_0+c_1\cdot (-q^2) +c_2\cdot(-iq_4) +c_3\cdot m_1 +c_4\cdot m_2.
\label{eq:qfit_sp}
\ee 
Solid line denotes $c_0+c_1\cdot (-q^2)$ with the fitting results
of $c_0$ and $c_1$.
We also draw $c_0+c_1\cdot (-q^2) +c_2\cdot(-i q_4) +c_3\cdot m_1 
+c_4\cdot m_2$ (dotted lines) choosing the four cases 
of $-iq_4=m_N-m_\pi$ with $m_1=m_2$.
The extrapolated value (open circle)
at the point $-q^2 a^2=m_N^2 a^2$ is consistent
with the result of $\alpha/(\sqrt{2}f)$ (triangle).
We observe similar situations for $\langle \pi^0|(ud_L) u_L|p\rangle$
and $\langle \pi^+|(ud_{R,L}) d_L|p\rangle$.

\section{Conclusions}
\label{sec:conclusion}

In this article we have reported progress in the lattice study of
the nucleon decay matrix elements.  
In order to enable a model-independent analysis of the nucleon decay, 
we have extracted the form factors of all the independent matrix elements
relevant for the (proton,neutron)$\rightarrow$($\pi,K,\eta$)
+(${\bar \nu},e^+,\mu^+$) decay processes without invoking
chiral lagrangian. 

We have also pointed out the necessity of separating out the 
contribution of an irrelevant form factor in lattice calculations 
for a correct estimate of the matrix elements at the physical point. 
With this separation, the matrix elements obtained from the three-point
functions are roughly 
comparable with the tree-level predictions of chiral lagrangian
with the $\alpha$ and $\beta$ parameters 
determined on the same lattice.
The magnitude of the matrix elements, however, are 
3 to 5 times larger than those with the smallest
estimate of $\alpha$ and $\beta$ among various QCD model
calculations.
Our results would stimulate phenomenological interests as
the larger values of the nucleon decay matrix elements
can give more stringent constraints on GUT models.

The ultimate goal of lattice QCD calculations 
of the nucleon decay matrix elements is to
determine the matrix elements precisely
with control over possible systematic errors.
Major systematic errors conceivably affecting our present results are 
the scaling violations and the quenching effects.
The former can be investigated by repeating the simulation
at several lattice spacings;
the latter is eliminated 
once configurations are generated with dynamical
quarks, where it is straightforward to apply our method. 
We leave these points to future studies.

\acknowledgements

One of us (Y.K.) thanks C.~Bernard for useful discussions.
This work is supported by the Supercomputer Project No.45 (FY1999)
of High Energy Accelerator Research Organization (KEK),
and also in part by the Grants-in-Aid of the Ministry of 
Education (Nos. 09304029, 10640246, 10640248, 10740107, 10740125,
11640294, 11740162).  K-I.I is supported by the JSPS Research
Fellowship.

\section*{appendix}

The perturbative renormalization factors for
the baryon number violating operators ${\cal O}_{RL}$ and ${\cal O}_{LL}$, 
which are defined in eqs.~(\ref{eq:z_rl}) and (\ref{eq:z_ll}), have already 
been calculated in Ref.~\cite{pt_w} 
employing the DRED scheme for the continuum theory. 
However, the authors of Ref.~\cite{pt_w} present only the numerical results 
for $Z$, $Z_{mix}$ and $Z^\prime_{mix}$. We consider that it 
would be instructive
to demonstrate the calculation of the renormalization factors
in detail.

We first rewrite the operators ${\cal O}_{RL}$ and ${\cal O}_{LL}$ as
\ben
{\cal O}_{RL}=\epsilon_{ijk}\left(({\bar \psi}^c_1)^i P_{R}(\psi_2)^j\right) 
P_L (\psi_3)^{k},
\label{eq:o_rl_c}\\
{\cal O}_{LL}=\epsilon_{ijk}\left(({\bar \psi}^c_1)^i P_{L}(\psi_2)^j\right) 
P_L (\psi_3)^{k},
\label{eq:o_ll_c}
\een
where ${\bar \psi}^c=\psi^T C$ is a charge conjugated field of $\psi$.
The continuum and Wilson quark actions for the charge conjugated 
field $\psi^c$ 
are obtained from those for $\psi$ with the replacement of
\be
igT^A\rightarrow -ig(T^A)^T,
\ee
where $T^A$ ($A=1,\dots,8$) are generators of color SU(3) group.
This implies the modification of the Feynman rule 
of the quark-gluon vertex for the $\psi^c$ field.

\begin{figure}[h]
\centering{
\hskip -0.0cm
\psfig{file=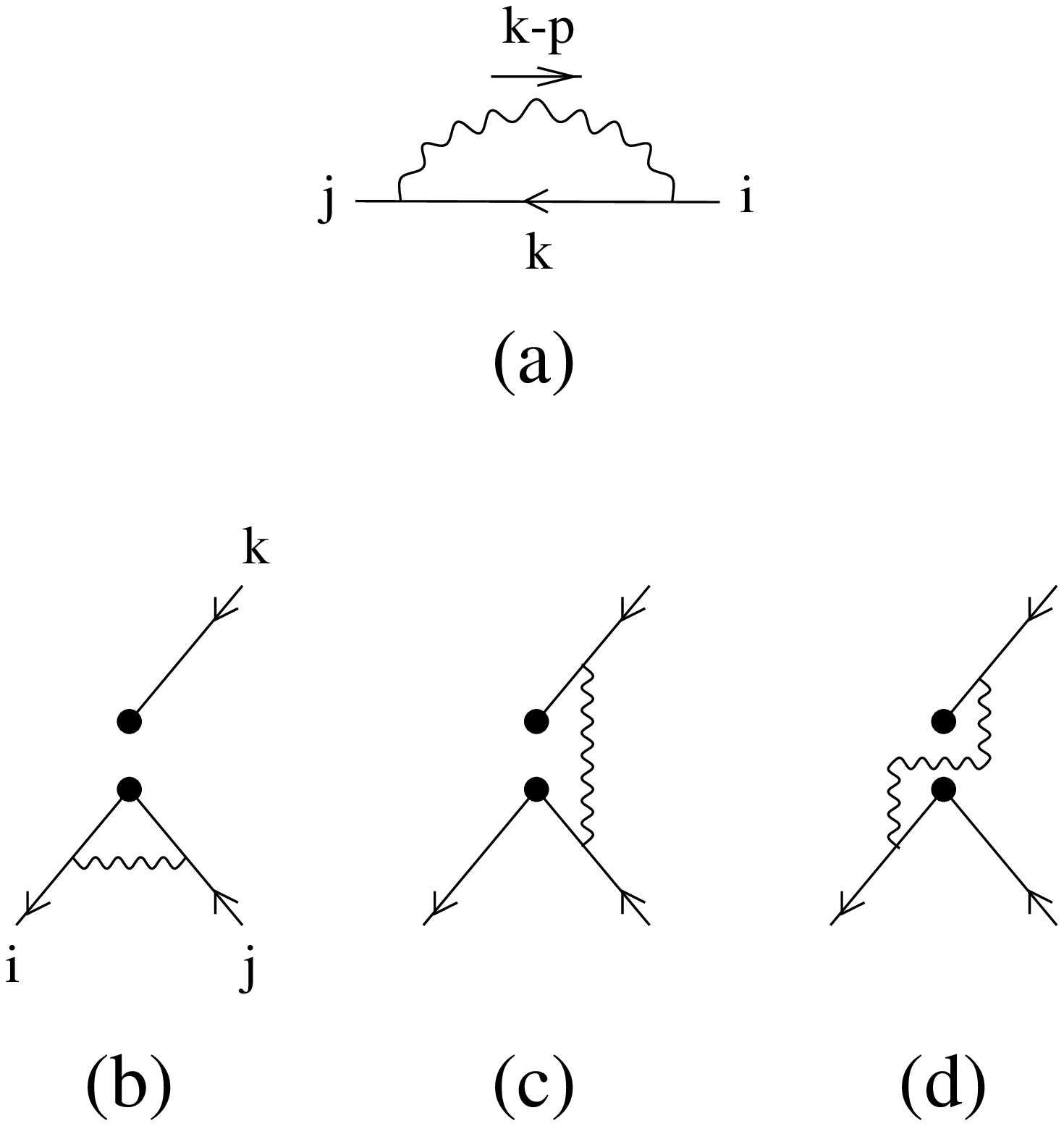,width=120mm,angle=-90}
}
\caption{One-loop diagrams for (a) quark self energy and (b)$-$(d)
vertex corrections for the three-quark operator. $p$ denotes a external
quark momentum and $i$, $j$ and $k$ at the ends of quark lines 
label color indices.} 
\label{fig:ptdgm}
\end{figure}

We illustrate the relevant one-loop diagrams in Fig.~\ref{fig:ptdgm}:
(a) the quark self energy and (b)-(d) the three types of 
vertex corrections. We calculate these diagrams in the Feynman gauge
for massless quarks with vanishing external momenta.
The infrared divergences are regularized by introducing
the fictitious small mass $\lambda$ in the gluon propagator:
\ben 
{G_{\mu\nu}^{ab}}^{\rm cont}&=&\delta_{ab}\delta_{\mu\nu}
\frac{1}{k^2+\lambda^2}, \\
{G_{\mu\nu}^{ab}}^{\rm latt}&=&\delta_{ab}\delta_{\mu\nu}
\frac{1}{4\sum_\alpha {\rm sin}^2(k_\alpha/2)+\lambda^2}.
\een
We should note that the infrared behavior of the theory should 
be independent of
the ultraviolet regularization schemes. The infrared divergent 
contributions in the one-loop diagrams, which emerges as the
${\rm ln}\lambda^2$ terms, are supposed to cancel in the
renormalization factors relating the continuum and lattice operators. 

Up to the one-loop level the inverse quark propagator and 
the vertex functions are written in the following form:
\ben
G^{-1}(p,\lambda)&=&i\slas{p}{}
\left(1-\frac{\alpha_s}{4\pi}\Sigma^{(1)}(\lambda)\right),\\
\Lambda_{RL,LL}(\lambda)&=&P_{R,L}\otimes P_L
+\frac{\alpha_s}{4\pi}\Lambda^{(1)}_{RL,LL}(\lambda),
\een
where
the superscript ($i$) refers to the $i$-th loop level.
$\Lambda^{(1)}_{RL,LL}$ represents the sum of contributions
from the three diagrams in Figs.~\ref{fig:ptdgm}(b)$-$(d).
The continuum results for $\Sigma^{(1)}$ and $\Lambda^{(1)}_{RL,LL}$ 
are given by
\ben
\Sigma^{(1)}(\lambda)&=&-\frac{4}{3}\left[\left(\frac{2}{\epsilon}-\gamma+
{\rm ln}|4\pi|\right)+{\rm ln}\left|\frac{\mu^2}{\lambda^2}\right|
-\frac{1}{2}\right],
\label{eq:sigma_ndr}\\
\Lambda^{(1)}_{RL,LL}(\lambda)&=&4{P_{R,L}\otimes P_L}
\left[\left(\frac{2}{\epsilon}-\gamma+
{\rm ln}|4\pi|\right)+{\rm ln}\left|\frac{\mu^2}{\lambda^2}\right|
+\frac{2}{3}\right],
\label{eq:lambda_ndr}
\een
in the NDR scheme  and
\ben
\Sigma^{(1)}(\lambda)&=&-\frac{4}{3}\left[\left(\frac{2}{\epsilon}-\gamma+
{\rm ln}|4\pi|\right)+{\rm ln}\left|\frac{\mu^2}{\lambda^2}\right|
+\frac{1}{2}\right],
\label{eq:sigma_dred}\\
\Lambda^{(1)}_{RL,LL}(\lambda)&=&4{P_{R,L}\otimes P_L}
\left[\left(\frac{2}{\epsilon}-\gamma+
{\rm ln}|4\pi|\right)+{\rm ln}\left|\frac{\mu^2}{\lambda^2}\right|+1\right],
\label{eq:lambda_dred}
\een
in the DRED scheme, where
the reduced space-time dimension $D$ is
parameterized by $\epsilon$ as $D=4-\epsilon$, $\epsilon > 0$.
The pole term $(2/\epsilon-\gamma+{\rm ln}|4\pi|)$ should be eliminated
in the ${\overline{\rm MS}}$ subtraction scheme.
The corresponding lattice results for  $\Sigma^{(1)}$ and 
$\Lambda^{(1)}_{RL,LL}$ are
\ben
\Sigma^{(1)}(\lambda)&=&-\frac{4}{3}
\left[{\rm ln}\left|\frac{\Lambda^2}{\lambda^2}\right|-1\right]\nn\\
&&+\frac{4}{3}(4\pi)^2\int_{-\pi}^\pi\frac{d^4 k}{(2\pi)^4}
\left[\frac{1}{\Delta_2(4\Delta_1+\lambda^2)}
\left(-\frac{1+r^2}{8}\Delta_4+r^2\Delta_1(2-\Delta_1)\right)\right.\nn\\
&&\left.+\frac{1}{\Delta_2(4\Delta_1+\lambda^2)^2}
\left(\frac{1+r^2}{2}\Delta_1\Delta_4-\Delta_4-\Delta_5\right)
+\frac{1}{2(4\Delta_1+\lambda^2)}\right.\nn\\
&&\left.\left.-\theta(\Lambda^2-k^2)\frac{-1}{(k^2+\lambda^2)^2}\right]
\right|_{\lambda=0}, 
\label{eq:sigma_latt}\\
\Lambda^{(1)}_{RL}(\lambda)&=&
4{\rm ln}\left|\frac{\Lambda^2}{\lambda^2}\right|\nn\\
&&+\frac{2}{3}(4\pi)^2\int_{-\pi}^\pi\frac{d^4 k}{(2\pi)^4}
\left[\frac{P_R\otimes P_L}{\Delta_2^2(4\Delta_1+\lambda^2)}
\left(\frac{\Delta_6}{2}(4r^2\Delta_1^2-\Delta_4)
+\frac{\Delta_2}{2}(4-(1-r^2)\Delta_1)\right.\right.\nn\\
&&\left.\left.+2\Delta_4-2\Delta_5+6r^2\Delta_1\Delta_4
+8r^4\Delta_1^3\right)\right.\nn\\
&&\left.+\frac{P_L\otimes P_L}{\Delta_2^2(4\Delta_1+\lambda^2)}
\left(\frac{\Delta_6}{2}(4r^2\Delta_1^2-\Delta_4)
-\frac{\Delta_2}{2}(4-(1-r^2)\Delta_1)
+2r^2\Delta_1\Delta_4\right)\right.\nn\\
&&\left.+\frac{\sum_\alpha\gamma_\alpha\gamma_5\otimes P_L\gamma_\alpha}
{\Delta_2^2(4\Delta_1+\lambda^2)}
\left(\frac{r^2}{4}\Delta_1\Delta_4-4r^2\Delta_1^2+r^2\Delta_1^3\right)
\right.\nn \\
&&\left.\left.-\theta(\Lambda^2-k^2)\frac{6}{k^2(k^2+\lambda^2)}\right]
\right|_{\lambda=0}, 
\label{eq:lambda_latt_rl}\\
\Lambda^{(1)}_{LL}(\lambda)&=&
4{\rm ln}\left|\frac{\Lambda^2}{\lambda^2}\right|\nn\\
&&+\frac{2}{3}(4\pi)^2\int_{-\pi}^\pi\frac{d^4 k}{(2\pi)^4}
\left[\frac{P_L\otimes P_L}{\Delta_2^2(4\Delta_1+\lambda^2)}
\left(\frac{\Delta_6}{2}(4r^2\Delta_1^2-\Delta_4)
+\frac{\Delta_2}{2}(4-(1-r^2)\Delta_1)\right.\right.\nn\\
&&\left.\left.+2\Delta_4-2\Delta_5+6r^2\Delta_1\Delta_4
+8r^4\Delta_1^3\right)\right.\nn\\
&&\left.+\frac{P_R\otimes P_L}{\Delta_2^2(4\Delta_1+\lambda^2)}
\left(\frac{\Delta_6}{2}(4r^2\Delta_1^2-\Delta_4)
-\frac{\Delta_2}{2}(4-(1-r^2)\Delta_1)
+2r^2\Delta_1\Delta_4\right)\right.\nn\\
&&\left.-\frac{\sum_\alpha\gamma_\alpha\gamma_5\otimes P_L\gamma_\alpha}
{\Delta_2^2(4\Delta_1+\lambda^2)}
\left(\frac{r^2}{4}\Delta_1\Delta_4-4r^2\Delta_1^2+r^2\Delta_1^3\right)
\right.\nn \\
&&\left.\left.-\theta(\Lambda^2-k^2)\frac{6}{k^2(k^2+\lambda^2)}\right]
\right|_{\lambda=0},
\label{eq:lambda_latt_ll} 
\een
where $r$ denotes the Wilson parameter and $\Delta_1$, $\Delta_2$,
$\Delta_4$, $\Delta_5$ and  $\Delta_6$ are given by
\ben
\Delta_1&=&\sum_\alpha {\rm sin}^2(k_\alpha/2), \\
\Delta_4&=&\sum_\alpha {\rm sin}^2(k_\alpha), \\
\Delta_5&=&\sum_\alpha {\rm sin}^2(k_\alpha){\rm sin}^2(k_\alpha/2), \\
\Delta_2&=&\Delta_4+4 r^2 \Delta_1^2, \\
\Delta_6&=&(1+r^2)\Delta_1-4.
\een
The counter terms in proportion to $\theta(\Lambda^2-k^2)$, 
which have the same infrared singularities as the lattice integrands, 
are introduced to pick out the analytical expressions of the
infrared divergent contributions.
The hyper-sphere radius $\Lambda$ does not exceed $\pi$.
With the use of eqs.~(\ref{eq:sigma_ndr})$-$(\ref{eq:lambda_dred}) 
and (\ref{eq:sigma_latt})$-$(\ref{eq:lambda_latt_ll}), we obtain 
the expression for the renormalization
constant $\Delta_{\slal{B}{}}$  in eq.~(\ref{eq:z_overall}), 
\ben 
\Delta_{\slal{B}{}}^{\rm NDR}&=& \Delta_{\slal{B}{}}^{\rm DRED}+\frac{2}{3}\nn\\
&=&\frac{5}{3}-4{\rm ln}|\Lambda|\nn\\
&&-(4\pi)^2\int_{-\pi}^\pi\frac{d^4 k}{(2\pi)^4}
\left[\frac{1}{2\Delta_1\Delta_2}
\left(-\frac{1+r^2}{8}\Delta_4+r^2\Delta_1(2-\Delta_1)\right)\right.\nn\\
&&\left.+\frac{1}{8\Delta_1^2\Delta_2}
\left(\frac{1+r^2}{2}\Delta_1\Delta_4-\Delta_4-\Delta_5\right)
+\frac{1}{4\Delta_1}\right.\nn\\
&&\left.+\frac{1}{6\Delta_1\Delta_2^2}
\left(\frac{\Delta_6}{2}(4r^2\Delta_1^2-\Delta_4)
+\frac{\Delta_2}{2}(4-(1-r^2)\Delta_1)\right.\right.\nn\\
&&\left.\left.+2\Delta_4-2\Delta_5+6r^2\Delta_1\Delta_4
+8r^4\Delta_1^3\right)\right.\nn\\
&&\left.-\theta(\Lambda^2-k^2)\frac{2}{k^4}\right].
\een
The mixing coefficients $Z_{mix}$ and $Z^\prime_{mix}$ 
in eqs.(\ref{eq:z_rl}) and (\ref{eq:z_ll}) are expressed by
\ben
Z_{mix}&=&(4\pi)^2\int_{-\pi}^\pi\frac{d^4 k}{(2\pi)^4}
\frac{-1}{6\Delta_1\Delta_2^2}
\left(\frac{\Delta_6}{2}(4r^2\Delta_1^2-\Delta_4)
-\frac{\Delta_2}{2}(4-(1-r^2)\Delta_1)
+2r^2\Delta_1\Delta_4\right),\\
Z_{mix}^\prime&=&(4\pi)^2\int_{-\pi}^\pi\frac{d^4 k}{(2\pi)^4}
\frac{1}{6\Delta_2^2}
\left(\frac{r^2}{4}\Delta_4-4r^2\Delta_1+r^2\Delta_1^2\right).
\een
We evaluate numerical values of $\Delta_{\slal{B}{}}^{\rm NDR}$,
$Z_{mix}$ and $Z_{mix}^\prime$ with $r=1$ using the Monte Carlo integration 
routine BASES\cite{bases}, which are already presented in
Sec.~\ref{subsec:calmethod}. Our results for $Z_{mix}$ and $Z_{mix}^\prime$
are consistent with those in Ref.~\cite{pt_w}, while
we observe a slight deviation beyond the statistical error
of the numerical integration for 
$\Delta_{\slal{B}{}}^{\rm DRED}$.

\newpage

\newpage

\newpage

\newpage

\newpage

\newpage

\newpage

\newpage

\newpage

\newpage
 
\newpage

\newpage

\newpage

\newpage
 
\newpage

\newpage

\newpage

\begin{table}[h]
\begin{center}
\caption{\label{tab:hmass}Hadron masses 
at $\beta=6.0$ in quenched QCD.}
\begin{tabular}{llll}
$K$ & $m_{\pi}a$ & $m_{\rho}a$ & $m_Na$  \\
\hline
0.15464      & 0.3209(8)  & 0.4350(22) & 0.6674(38)  \\
0.15516      & 0.2843(9)  & 0.4135(26) & 0.6253(43)  \\
0.15568      & 0.2436(9)  & 0.3925(33) & 0.5814(52)  \\
0.15620      & 0.1957(11) & 0.3723(45) & 0.5359(69)  \\
0.157136(12) &            & 0.3346(56) & 0.4607(89)  \\
\end{tabular}
\end{center}
\end{table}

\begin{table}[h]
\begin{center}
\caption{\label{tab:q2a2}Four-momentum transfers
from the nucleon at rest to the pseudoscalar meson.
$E_{PS}$ is the energy of the pseudoscalar meson with 
spatial momentum ${\vec p}$.}
\begin{tabular}{lllrllr}
 & & {$|\vec{p}|a=0$}
   & \multicolumn{2}{c}{$|\vec{p}|a=\pi/24$} 
   & \multicolumn{2}{c}{$|\vec{p}|a=\pi/14$} \\
$K$ for $m_1$ & $K$ for $m_2$ & $-q^2a^2$ & $E_{PS}\cdot a$ & $-q^2a^2$ 
                          & $E_{PS}\cdot a$ & $-q^2a^2$ \\
\hline
0.15464  & 0.15464 & 0.1200(28)   & 0.3467(10)  & 0.0857(26) & 0.3912(12)  & $ 0.0259(22)$ \\
         & 0.15516 && 0.3302(11)  & 0.0966(27) & 0.3767(13)  & $ 0.0341(24)$ \\
         & 0.15568 && 0.3131(12)  & 0.1084(29) & 0.3619(15)  & $ 0.0429(25)$ \\
         & 0.15620 && 0.2953(14)  & 0.1213(31) & 0.3469(21)  & $ 0.0523(28)$ \\
0.15516  & 0.15464 && 0.3304(10)  & 0.0698(27) & 0.3769(13)  & $ 0.0113(23)$ \\
         & 0.15516 & 0.1163(31)   & 0.3132(11) & 0.0803(29)  & 0.3620(14)  & $ 0.0190(25)$ \\
         & 0.15568 && 0.2953(13)  & 0.0918(30) & 0.3469(15)  & $ 0.0272(26)$ \\
         & 0.15620 && 0.2765(15)  & 0.1045(32) & 0.3316(22)  & $ 0.0359(28)$ \\
0.15568  & 0.15464 && 0.3138(11)  & 0.0545(29) & 0.3626(14)  & $-0.0025(24)$ \\
         & 0.15516 && 0.2957(12)  & 0.0645(31) & 0.3473(16)  & $ 0.0045(26)$ \\
         & 0.15568 & 0.1141(36)   & 0.2768(14) & 0.0757(33)  & 0.3317(19)  & $ 0.0120(28)$ \\
         & 0.15620 && 0.2567(16)  & 0.0883(35) & 0.3159(25)  & $ 0.0202(31)$ \\
0.15620  & 0.15464 && 0.2969(13)  & 0.0400(33) & 0.3480(18)  & $-0.0150(27)$ \\
         & 0.15516 && 0.2777(14)  & 0.0495(36) & 0.3322(19)  & $-0.0089(29)$ \\
         & 0.15568 && 0.2575(15)  & 0.0604(39) & 0.3163(23)  & $-0.0021(32)$ \\
         & 0.15620 & 0.1158(47)   & 0.2356(18) & 0.0730(42)  & 0.3003(31)  & $ 0.0052(37)$ \\
\end{tabular}
\end{center}
\end{table}

\begin{table}[h]
\begin{center}
\caption{\label{tab:ab}Results for $\alpha$ and $\beta$ parameters
as a function of quark mass.
Operators are renormalized at the scale $\mu$ in the NDR scheme.}
\begin{tabular}{lllll}
    & \multicolumn{2}{c}{$\alpha a^3$} & \multicolumn{2}{c}{$\beta a^3$} \\
$K$ & {$\mu=1/a$} & {$\mu=\pi/a$} & {$\mu=1/a$} & {$\mu=\pi/a$} \\
\hline
0.15464      & $-0.00180(6)$  & $-0.00207(7)$  & 0.00179(6)  & 0.00205(7)  \\
0.15516      & $-0.00166(7)$  & $-0.00191(8)$  & 0.00165(7)  & 0.00189(8)  \\
0.15568      & $-0.00155(7)$  & $-0.00177(8)$  & 0.00152(7)  & 0.00174(8)  \\
0.15620      & $-0.00148(9)$  & $-0.00170(11)$ & 0.00143(9)  & 0.00164(10) \\
0.157136(12) & $-0.00125(11)$ & $-0.00144(13)$ & 0.00119(11) & 0.00137(12) \\
\end{tabular}
\end{center}
\end{table}

\begin{table}[h]
\begin{center}
\caption{\label{tab:ab_latt} Comparison of $\alpha$ and $\beta$ parameters
in lattice QCD. 
All calculations are done with the Wilson quark action 
in the quenched approximation.
Lattice cutoff $a^{-1}$ is determined from $m_\rho$.
Quark mass is defined by 
$m_q=\left(1/2K-1/2K_c\right)\cdot a^{-1}$.}
\begin{tabular}{ccccccccc}
Ref. & Hara {\it et al.} \cite{hara,hara_spc} 
     & Bowler {\it et al.} \cite{bowler,bowler_spc} 
     & Gavela {\it et al.} \cite{gavela,ms_spc}
     & This work \\
\hline
Lattice size & $16^3\times 48$ & $8^3\times 16$ 
             & $10^2\times20\times 40$ & $28^2\times48\times 80$ \\
No. config.   & 15 & 32 & 30 & 100 \\
$a^{-1}$ [GeV] & 1.81(6) & 1.45(9) & $2.2(4)$ & $2.30(4)$ \\
Spatial size [fm$^{3}$] & $(1.7)^3$ & $(1.1)^3$ 
                        & $(0.9)^2\times 1.8$ & $(2.4)^2\times 4.1$ \\   
Quark mass [MeV] & $109 \simlt m_q \simlt 696$ 
                 & $184 \simlt m_q \simlt 477$
                 & $82 \simlt m_q \simlt 223$
                 & $44 \simlt m_q \simlt 118$ \\
$\alpha$ [GeV$^3$]  &                     & $|\alpha|\sim 0.065$
                    & $|\alpha|=0.019(2)$ & $\alpha=-0.015(1)$ \\
$\beta$  [GeV$^3$]  & $|\beta|=0.029(6)$  & $|\beta|\sim 0.050$
                    &                     & $\beta=0.014(1)$ \\
Renorm. scheme & & Pauli-Villars & DRED & NDR \\
Renorm. scale  & & $\mu=85$ GeV & $\mu=1/a$ & $\mu=1/a$ \\
\end{tabular}
\end{center}
\end{table}

\begin{table}[h]
\begin{center}
\caption{\label{tab:summary}Results for relevant form factors
in the independent nucleon decay matrix elements 
of eqs.~(\protect{\ref{eq:indme_1}})$-$(\protect{\ref{eq:indme_7}}).
Operators are renormalized at the scale $\mu$ in the NDR scheme.}
\begin{tabular}{lrrrr}
matrix element & \multicolumn{2}{c}{$\mu=1/a$} & \multicolumn{2}{c}{$\mu=\pi/a$} \\
                & lattice units & GeV$^2$ & lattice units & GeV$^2$ \\
\hline
$\la \pi^0  | (u d_R)u_L | p \ra$ & $-0.0253(31)$ & $-0.134(16)$  & $-0.0289(36)$ & $-0.153(19)$ \\
$\la \pi^0  | (u d_L)u_L | p \ra$ & $ 0.0242(32)$ & $ 0.128(17)$  & $ 0.0278(37)$ & $ 0.147(20)$ \\
$\la \pi^+  | (u d_R)d_L | p \ra$ & $-0.0357(45)$ & $-0.189(24)$  & $-0.0409(51)$ & $-0.216(27)$ \\
$\la \pi^+  | (u d_L)d_L | p \ra$ & $ 0.0343(45)$ & $ 0.181(24)$  & $ 0.0394(52)$ & $ 0.208(28)$ \\
$\la K^0    | (u s_R)u_L | p \ra$ & $ 0.0192(20)$ & $ 0.102(11)$  & $ 0.0213(22)$ & $ 0.113(12)$ \\
$\la K^0    | (u s_L)u_L | p \ra$ & $ 0.0089(12)$ & $ 0.0471(63)$ & $ 0.0092(13)$ & $ 0.0487(69)$ \\
$\la K^+    | (u s_R)d_L | p \ra$ & $-0.0100(13)$ & $-0.0529(69)$ & $-0.0118(15)$ & $-0.0624(79)$ \\
$\la K^+    | (u s_L)d_L | p \ra$ & $ 0.0088(11)$ & $ 0.0466(58)$ & $ 0.0105(13)$ & $ 0.0555(69)$ \\
$\la K^+    | (u d_R)s_L | p \ra$ & $-0.0248(27)$ & $-0.131(14)$  & $-0.0282(31)$ & $-0.149(16)$ \\
$\la K^+    | (u d_L)s_L | p \ra$ & $ 0.0268(28)$ & $ 0.142(15)$  & $ 0.0304(32)$ & $ 0.161(17)$ \\
$\la K^0    | (u s_R)d_L | n \ra$ & $ 0.0090(12)$ & $ 0.0476(63)$ & $ 0.0094(13)$ & $ 0.0497(69)$ \\
$\la K^0    | (u s_L)d_L | n \ra$ & $ 0.0179(18)$ & $ 0.0947(95)$ & $ 0.0198(20)$ & $ 0.105(11)$ \\
$\la \eta   | (u d_R)u_L | p \ra$ & $ 0.0007(11)$ & $ 0.0037(58)$ & $ 0.0004(12)$ & $ 0.0021(63)$ \\
$\la \eta   | (u d_L)u_L | p \ra$ & $ 0.0191(25)$ & $ 0.101(13)$  & $ 0.0212(28)$ & $ 0.112(15)$ \\
\end{tabular}
\end{center}
\end{table}

\end{document}